\pdfoutput=1
\documentclass[aps,prd,twocolumn, floatfix, nofootinbib,longbibliography]{revtex4-2}
\usepackage[utf8]{inputenc}
\usepackage[english]{babel}
\usepackage{natbib}
\usepackage{graphicx}
\usepackage{amsmath, amssymb, amsthm}
\usepackage{tensor}
\usepackage{braket}
\usepackage{cases}
\usepackage{dsfont}
\usepackage{hyperref}
\usepackage{svg}
\usepackage[export]{adjustbox}
\usepackage{leftidx}
\usepackage{BOONDOX-cal} 
\usepackage{bbm} 
\usepackage{faktor}
\usepackage[percent]{overpic}
\usepackage{adjustbox}
\usepackage{pifont}
\usepackage{enumitem}
\usepackage{epigraph}
\usepackage{xcolor}
\usepackage{orcidlink}

\numberwithin{equation}{section}

\DeclareMathOperator{\Diff}{Diff}
\DeclareMathOperator{\TDiff}{TDiff}
\DeclareMathOperator{\GS}{GS}
\DeclareMathOperator{\T}{T}
\DeclareMathOperator{\Spann}{span}
\DeclareMathOperator{\Cyl}{Cyl}

\newcommand{\localtextbulletone}{\textcolor{black}{\raisebox{.45ex}{\rule{.75ex}{.75ex}}}}

\newcommand{\comm}[2]{\left[ #1 , #2 \right]}

\newcommand{\vac}{\left\lvert 0 \right\rangle}

\newcommand{\jrep}[2]{\pi_{#1}\left({#2}\right)}

\newcommand{\scpr}[2]{\langle#1\, \vert \, #2 \rangle}

\newcommand{\gbra}[1]{(\,#1\,\rvert}
\newcommand{\gscpr}[2]{(\,#1\, \vert \, #2\, \rangle}

\newcommand{\elmagA}[0]{\underline{A}}
\newcommand{\elmagE}[0]{\underline{E}}
\newcommand{\elmagB}[0]{\underline{B}}
\newcommand{\elmagF}[0]{\underline{F}}
\newcommand{\elmagh}[1]{\underline{h}_{#1}}

\newtheorem{definition}{Definition}[section]
\newtheorem{lemma}{Lemma}[section]

\newtheorem{example}[definition]{Example}

\begin{document}
\title{Kinematics of Arbitrary Spin Matter Fields in Loop Quantum Gravity}

\author{Refik Mansuroglu \orcidlink{0000-0001-7352-513X}}
\email[]{Refik.Mansuroglu@fau.de}
\affiliation{Institute for Quantum Gravity, Friedrich-Alexander-Universität Erlangen-Nürnberg (FAU), Staudtstraße 7, 91058 Erlangen, Germany}
\author{Hanno Sahlmann \orcidlink{0000-0002-8083-7139}}
\email[]{Hanno.Sahlmann@gravity.fau.de}
\affiliation{Institute for Quantum Gravity, Friedrich-Alexander-Universität Erlangen-Nürnberg (FAU), Staudtstraße 7, 91058 Erlangen, Germany}

\date{\today}

\begin{abstract}
\noindent Loop quantum gravity envisions a small scale structure of spacetime that is markedly different from that of the classical spacetime continuum. This has ramifications for the excitation of matter fields and for their coupling to gravity. There is a general understanding of how to formulate scalar fields, spin $\frac{1}{2}$ fields and gauge fields in the framework of loop quantum gravity. The goal of the present work is 
to investigate kinematical aspects of this coupling. 

We will study implications of the Gauß and diffeomorphism constraint for the quantum theory: We define and study a less ambiguous variant of the Baez-Krasnov path observables, and investigate symmetry properties of spin network states imposed by diffeomorphism group averaging. We will do this in a setting which allows for matter excitations of spin $\frac{1}{2}$ and higher. In the case of spin $\frac{1}{2}$, we will also discuss extensions of it by introducing an electromagnetic field and antiparticles.
We finally discuss in how far the picture with matter excitations of higher spin can be obtained from classical actions for higher spin fields. 
\end{abstract}

\maketitle

\section{Introduction}
The understanding of quantum matter fields combined with a theory of quantum gravity is an important step towards a grand unified theory. On the one hand, matter fields yield access to verifying the theory of quantum gravity. On the other hand, quantum gravity can act as a natural regulator of quantum matter which solves conceptual problems in quantum field theory.

Loop quantum gravity uses a formulation of general relativity as a constrained gauge theory, with a Gau{\ss} constraint encoding SU(2) gauge invariance (invariance under spatial frame rotations). It was realized early on \cite{MoralesTecotl:1994ns, MoralesTecotl:1995jh, Baez:1997bw} that, to solve the Gauss constraint, gravity and fermionic excitations have to be coupled. A very compelling solution was first suggested in \cite{MoralesTecotl:1994ns, MoralesTecotl:1995jh} and later expanded on in \cite{Baez:1997bw}: the fermions sit at the open ends of gravitational spin networks. It was then realized that to consistently deal with adjointness relations, the density weight between the fermionic canonical variables has to be balanced \cite{Thiemann_1998}. Detailed derivations from classical actions have been considered \cite{Thiemann_1998_QSD,Bojowald:2007nu,Bojowald:2008nz}. More recently, the coupling of fermion states to gravity has been investigated also from the perspective of spin foam models for loop quantum gravity \cite{Han_2013, Bianchi_2013}. The coupled gravity-fermion states we are considering here are precisely the boundary states in the spinfoam formalism. 

In the present work we expand on this in two ways. On the one hand, the picture of spin $\frac{1}{2}$ matter immediately suggests a generalization to point excitations of spin other than $\frac{1}{2}$. We will use this general picture in most of the work, and also begin a discussion of how it could be derived from classical actions for fields of higher spin. On the other hand, we investigate the consequences of Gau{\ss} and diffeomorphism constraint, by studying various examples of quantum states and by general considerations.

In section \ref{sec:LQG_matter} we generalize the matter Hilbert space of \cite{Thiemann_1998} to excitations with arbitrary spin quantum numbers and use well-known methods to implement gauge invariance \cite{MoralesTecotl:1994ns, MoralesTecotl:1995jh, Baez:1997bw}. In particular, we combine the idea of the gauge invariant path observables of \cite{Baez:1997bw} with the quantum theory of \cite{Thiemann_1998} to obtain simpler gauge invariant observables.\\ 
In section \ref{sec:diffeo} we discuss the implementation  \cite{Ashtekar_1995, Giulini:1998rk, Giulini:1999kc} of symmetry under spatial diffeomorphisms in detail. For the case of spin $\frac{1}{2}$ we make sure that we can remove gauge transformations from the diffeomorphism constraint locally, to obtain a constraint that generates exclusively local spatial diffeomorphisms. 
We then discuss symmetry properties of the quantum states imposed by the diffeomorphism constraint and the statistics of the matter fields in examples. It turns out that assuming the spin statistics connection from quantum field theory, simple rules can be formulated for certain symmetric states to vanish with the implementation of the diffeomorphism constraint.\\
In section \ref{sec:actions} we suggest candidates for the classical actions describing the semiclassical limits of the considered quantum theory. A Hamiltonian formulation yields contributions to Gau{\ss}, diffeomorphism and Hamilton constraint constraints from the matter action, but also new constraints. We make some simple observations about this constrained Hamiltonian formulation, but also point out thorny issues that makes those classical theories quite complicated.\\  
In section \ref{sec:elmag}, an embedding of a U(1) gauge symmetry into the theory for Dirac fermions is considered. The theory of electromagnetism is already well understood within the context of vacuum loop quantum gravity \cite{Corichi97}, and the coupling to fermions is contained in \cite{Thiemann_1998_QSD}. Our discussion leads to a formulation of positive and negative electromagnetic charges or particles and antiparticles, respectively.\\

Throughout the paper, we use the signature $(-+++)$ for the metric. The spatial slice the canonical theory will be based on is denoted by $\Sigma$. Four-dimensional spacetime indices are denoted by lower case Greek letters $\mu, \nu, \rho, ... \in \{0, ..., 3\}$. Spatial indices are denoted by lower case letters $a, b, c, ... \in \{1, 2, 3\}$.

Indices which correspond to a spin $\frac{1}{2}$ representation of SU(2) are denoted by capital letters $A, B, C \in \{1, 2\}$. Spin 1 representations are denoted by lower case letters starting with $i, j, k, ... \in \{1, 2, 3\}$ and four-dimensional spin $\frac{1}{2}$ (Dirac) representations are denoted by capital letters starting with $I, J, K, ... \in \{0, 1, 2, 3\}$. Higher spin representations are built by the symmetrized direct sum of Dirac representations and are indexed by the multiindices denoted by capital script letters $\mathcal{A}, \mathcal{B}, \mathcal{C}, ... \in \{ A, i, (A_1 A_2 A_3), ...  \}$.

\section{Loop Quantum Gravity with Matter Fields}
\label{sec:LQG_matter}
In this section, we will sketch the construction of an unconstrained Hilbert space for gravity and matter of arbitrary spin, and the implementation of the Gauss constraint. This is a natural generalization of the construction of \cite{Thiemann_1998}. We also consider the generalization of natural observables first suggested in \cite{MoralesTecotl:1994ns, MoralesTecotl:1995jh} and later studied in \cite{Baez:1997bw}. Using creation and annihilation operators, which both act pointwise, enables us to reduce an ambiguity of these \textit{path observables} \cite{Baez:1997bw}.

Let us start with matter-free loop quantum gravity. The gravitational observables act on cylindrical functions which form the Ashtekar-Lewandowski Hilbert space,
\begin{equation}
	\mathcal{H}_\text{AL}=L^2(\overline{\mathcal{A}},\text{d}\mu_{\text{AL}}),
\end{equation}
via multiplication of holonomies and the action of the derivation $X_S$, respectively \cite{Ashtekar:1996eg}
\begin{align}
	\label{eq:Grep}
	\jrep{j}{h}_e \Psi[A] &= \jrep{j}{h}_e[A] \Psi[A], \\
	\int_S E_i \Psi[A] &= i (X_S^i\Psi)[A].
\end{align}
The matter degrees of freedom are described by a Fock space based on point-like excitations. The total unconstrained Hilbert space can hence be described by the tensor product
\begin{align}
	\mathcal{H} = \mathcal{H_{\text{AL}}} \otimes \mathcal{F}^\pm \left( \mathcal{h}^{(j)} \right),
\end{align}
where $\mathcal{F}^\pm$ denotes the (anti)symmetric Fock space over the one particle Hilbert space 
\begin{equation}
	\mathcal{h}^{(j)}=\bigoplus_{x\in\Sigma} \mathbb{C}^{2j + 1}.   
\end{equation}
This can be equivalently written as
\begin{align}
	\mathcal{h}^{(j)} &= \{f:\Sigma \longrightarrow\mathbb{C}^{2j + 1}\rvert f(x)\neq 0 \text{ for finitely many } x \}\\
	\scpr{f}{f'} &= \sum_{x\in\Sigma}\overline{f(x)} f'(x).
\end{align}
The matter Fock space comes with creation and annihilation operators satisfying the canonical (anti)commutation relations,
\begin{align}
	\comm{\theta(x)^\mathcal{A}}{\theta^\dagger(y)_\mathcal{B}}_\pm &= \delta_\mathcal{B}^\mathcal{A} \delta_{x,y} \\
	\comm{\theta(x)^\mathcal{A}}{\theta(y)^\mathcal{B}}_\pm &= 0 \\
	\comm{\theta^\dagger (x)_\mathcal{A}}{\theta^\dagger(y)_\mathcal{B}}_\pm &= 0,
\end{align}
where we denoted the indices ranging over the spin $j$ representation space by the script letters $\mathcal{A}, \mathcal{B}$. These can be constructed by $2j$ many symmetrized Weyl spinor indices. $x,y$ are points in the spatial slice $\Sigma$. Note that the creation and annihilation operators can be constructed to both act as a spacetime scalar and hence pointwise \cite{Thiemann_1998}.

On the way towards a physical Hilbert space, we need to implement the Gauss constraint. For this, we only regard quantum states lying in the kernel of the quantum Gauss constraint operator, or equivalently states that are invariant with respect to the unitary action $U_g$ of gauge transformations generated by the Gauss constraint operator. The action of $U_g$ reads\footnote{Note that we are using the convention that the first index of $h_e$ transforms at $s(e)$ and the second one at $t(e)$. At the same time, we are using the convention for $e \circ f$, which is used in \cite{thiemann_2007}, such that $h_e \cdot h_f = h_{e \circ f}$ holds.}
\begin{align}
	U_g\jrep{j}{h_e} U^{-1}_g &= g(s(e)) \cdot\jrep{j}{h_e} \cdot g^{-1}(t(e)), \label{eq:trans_holonomy} \\
	U_g \theta(x) U^{-1}_g &= g(x) \cdot \theta(x), \\ 
	U_g \theta^\dagger (x) U^{-1}_g &= \theta^\dagger (x) \cdot g^{-1}(x). \label{eq:trans_matter}
\end{align}
When implementing the Gauss constraint, we arrive at the Hilbert space of SU(2) invariant states. One suitable basis is given by a generalization of spin network states, which also admit spin representations of matter fields at the vertices of the underlying spin network graph. We want to characterize these in the following subsection.

\subsection{Generalized Spin Network States}
The kinematic Hilbert space is spanned by the tensor products of spin network states with Fock states. To implement the Gauss constraint, one can follow either refined algebraic quantization \cite{Giulini:1998rk, Giulini_1999} or a reduced phase space quantization both yielding the same result.

We have to filter for all states which are invariant under gauge transformations \cite{MoralesTecotl:1994ns, Thiemann_1998_QSD} leaving only certain combinations of holonomies, intertwiners and matter fields where the gauge transforamtion cancels. Given a graph $\gamma$ and a set of matter fields $\theta_{p_1}, ..., \theta_{p_N}$, together with (\ref{eq:trans_holonomy} - \ref{eq:trans_matter}) we can now deduce the characteristics of the quantum states in the Hilbert space $\mathcal{H}_G$ of gauge invariant states with the following characteristics:
\begin{itemize}[label= \localtextbulletone]
	\item The matter field $\theta$ has to be attached to a vertex of the underlying graph $\gamma$. This might also be a 2-valent vertex, although we do not consider them in the vacuum theory.
	\item For an n-valent vertex with a single matter field $\theta$ of spin $j$ attached, only intertwiners of the form
	\begin{align}
		\iota: j_1 \otimes ... \otimes j_n \to j
	\end{align}
	can be gauge-invariantly coupled with $\theta$. In particular, the set of spin quantum numbers $(j_1, ..., j_n)$ is restricted by the Clebsch-Gordan rules for spin coupling.
	\item For an n-valent vertex with an arbitrary number $N$ of particles, only intertwiners of the form
	\begin{align}
		\iota: j_1 \otimes ... \otimes j_n \to k
	\end{align}
	can be gauge-invariantly coupled. Again, the set of spin quantum numbers $(j_1, ..., j_n)$ is restricted by the theory of spin coupling. Here $k$ denotes a spin quantum number the $N$ particles can couple to. $k$ is restricted by Clebsch-Gordan theory to be in the set \\
	\begin{align}
		&k \in \left\{0, 1, ..., N \cdot j \right\} \\
		\text{ or } &k \in \left\{ \frac{1}{2}, \frac{3}{2}, ..., N \cdot j \right\}.
	\end{align}
	The second set is valid only for $N$ odd and $j$ half-integral. If we assume anti-commutation relations, as known as fermionic quantization, then the total spin is further bounded by a total number of $2j + 1$ particles or equivalently $k \leq j + \frac{1}{2} - \frac{N}{2}$. This encodes the finiteness of the antisymmetrized Fock space at any point.
	
	In particular, $N = 0$ reduces to the vacuum spin network case. Furthermore, we can generalize this for mixing different types (spins) of particles at one and the same point. Then, we would get the tensor product of coupled spin variables $k_1, ..., k_l$ each corresponding to one specific type. The intertwiner then has the form
	\begin{align}
		\iota: j_1 \otimes ... \otimes j_n \to k_1 \otimes ... \otimes k_l.
	\end{align}
	\item 1-valent vertices, i.e. single starting- or endpoints of holonomies, are compatible with matter fields attached. As the corresponding intertwiner has to couple to 0, only a spin $k$ representation of the said holonomy qualifies for a gauge invariant spin network state.
\end{itemize}
The total Hilbert space of gauge invariant states is then spanned by the states described above. Another way to describe the gauge invariant states is that they are obtained by a pairing (summation over the free indices) of a generalized spin network as defined in \cite{Ashtekar:1996eg} with a suitable matter state with particle excitations at the non-gauge invariant vertices. 

\begin{figure}[t]
	\centering
	\includegraphics[scale=0.1]{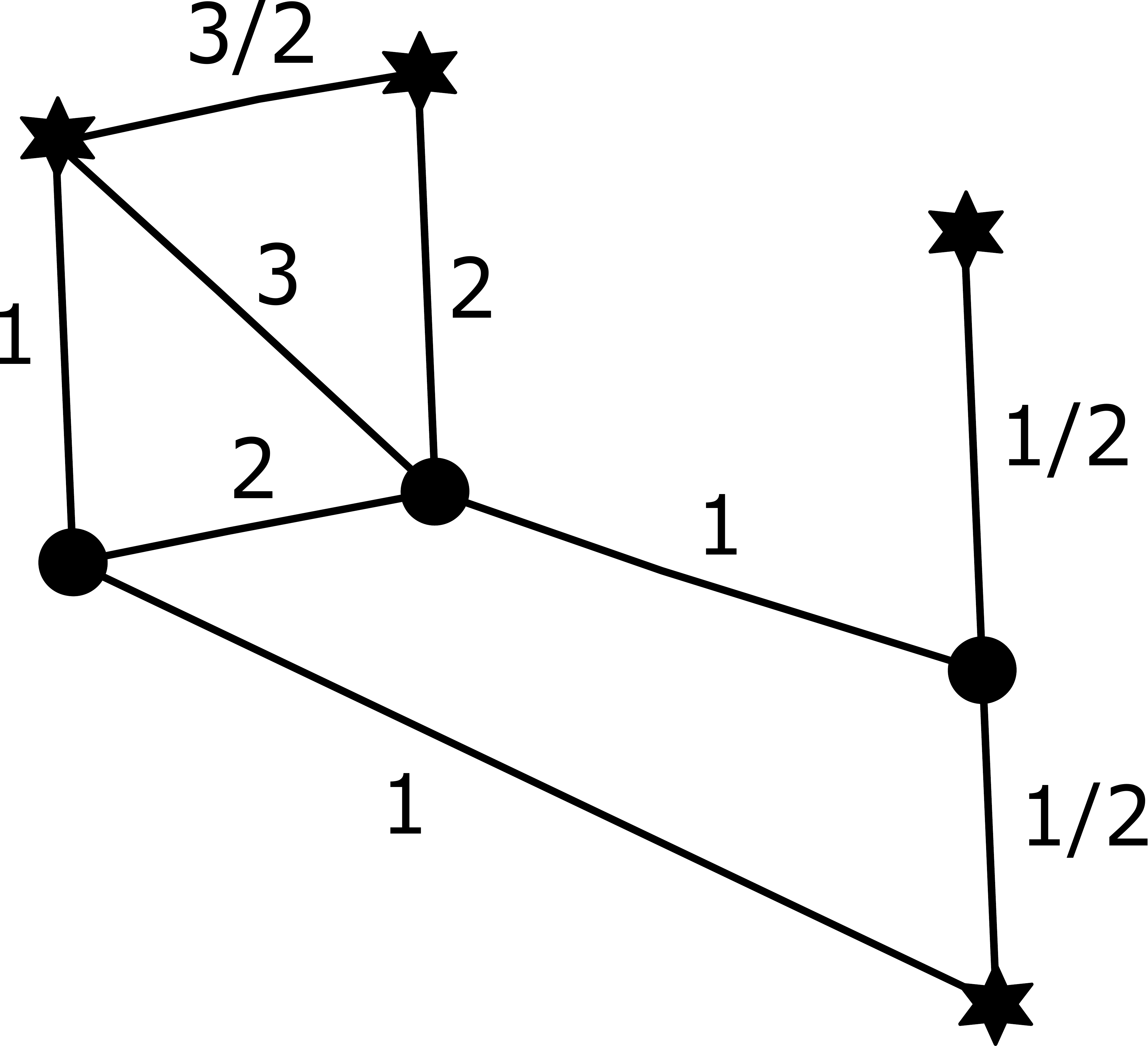}
	\caption{Exemplary generalized spin network state. Vertices with a spin $\frac{1}{2}$ particle are depicted by stars and vertices without matter by dots. The spin $\frac{1}{2}$ particles couple to the star intertwiner gauge invariantly. For the sake of clarity, arrows depicting the direction of the edges are omitted.}
	\label{fig:spin_network_spin1/2}
\end{figure}

We complete the discussion on the implementation of the Gauss constraint by discussing an example of such a \textit{generalized spin network state} shown in \autoref{fig:spin_network_spin1/2}.
There, we can see a graph with spin $\frac{1}{2}$ particles denoted by stars, which lie at the vertices. The black dots on the other hand denote vertices without matter. We might have also sketched a pair of spin $\frac{1}{2}$ particles, but they will have no effect to the state. Single particles do influence the spin quantum numbers of the adjacent holonomies as they have to couple to $j = \frac{1}{2}$ in order to yield a gauge invariant state in total. The generalized spin networks can be formulated with arbitrary particle types with an adequate pictorial notation. To keep the example simple, we leave it by inserting spin $\frac{1}{2}$ particles only. Furthermore note that we did not depict the directions of the edges in \autoref{fig:spin_network_spin1/2} for the sake of clarity. The direction can indicate index positions of holonomies and intertwiners. However, using the respective (pseudo-)metric, we can arbitrarily lower and raise the indices and therefore change the directions of the edges, anyway.

\subsection{Path Observables}
After we have introduced the Hilbert space $\mathcal{H}_G$ of gauge invariant generalized spin network states, we will now face the natural question of how to create and annihilate particles. Obviously, it is not possible to create or annihilate a single spin $\frac{1}{2}$ particle without leaving $\mathcal{H}_G$. Instead, we can introduce operators which are coupled gauge invariantly.

\cite{MoralesTecotl:1994ns, MoralesTecotl:1995jh} suggests to couple creators and annihilators by holonomies. The idea was continued by \cite{Baez:1997bw}. However there, the annihilation operator is a density of weight one. Therefore, every appearing annihilation operator has to be smeared gauge covariantly, i.e. with a holonomy with variable endpoint lying within an open subset $\mathcal{R} \subset \Sigma$ whose closure is compact. To do this in a well-defined manner, an arbitrary but fixed rule of how to construct the edge $e_{p p'}$ and therefore the holonomy $h_{e_{p p'}}$ with fixed starting point $p$ and variable endpoint $p'$ has to be applied. This is necessary since there are a priori infinitely many different edges with specific starting- and endpoints when integrating over the endpoint of the holonomy $h_{e_{p p'}}$.

In our case, using the scalar creation and annihilation operators of \cite{Thiemann_1998}, this complication is bypassed. A gauge invariant creation operator of two different chiral components may take the form
\begin{align}
	\hat{\theta}^A_p \epsilon_{AC} \tensor{\jrep{\frac{1}{2}}{h_e}}{^C_B} \hat{\theta}^B_q = \,\, \overset{j = \frac{1}{2}}{\leftidx{_p^\theta}{\includegraphics[valign=c, scale=0.12]{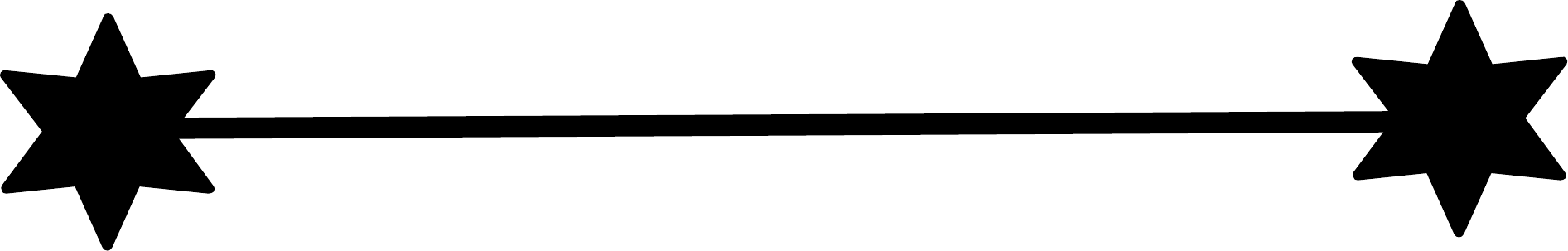}}{_q^\theta}}_,
	\label{eq:Baez_theta_theta}
\end{align}
where $p, q \in \Sigma$ and $e(0) = p, e(1)= q$ we depicted this so called \textit{path observable} also by a graph with the stars indicating creation of, in this case, two Weyl spinors. The direction of $e$ is again implicitly encoded in the starting and endpoint $p, q$. The order of the action of the operators follows the convention that the symbol on the right acts first (just as in the algebraic formulae).

The operator (\ref{eq:Baez_theta_theta}) acts on a generalized spin network state by generating a matter field at the point $p$, a spin $\frac{1}{2}$ holonomy along the edge $e$ and a matter field at the point $q$. The resulting state can be written again as a linear combination of spin network states now corresponding to a potentially larger graph.

If we wanted to define an operator which annihilates a particle at one or both endpoints of the holonomy, we have to be careful whether one uses already the smeared version of the annihilation operator or the version of \cite{MoralesTecotl:1994ns} of the annihilation operator. With the smeared version $\hat{\theta}^\dagger_A$, we can define, for instance
\begin{align}
	\hat{\theta}^\dagger_{p \, A} \tensor{\jrep{\frac{1}{2}}{h_e}}{^A_B} \hat{\theta}^B_q = \,\, \stackrel{j = \frac{1}{2}}{\leftidx{_p^{\theta^\dagger}}{\includegraphics[valign=c, scale=0.12]{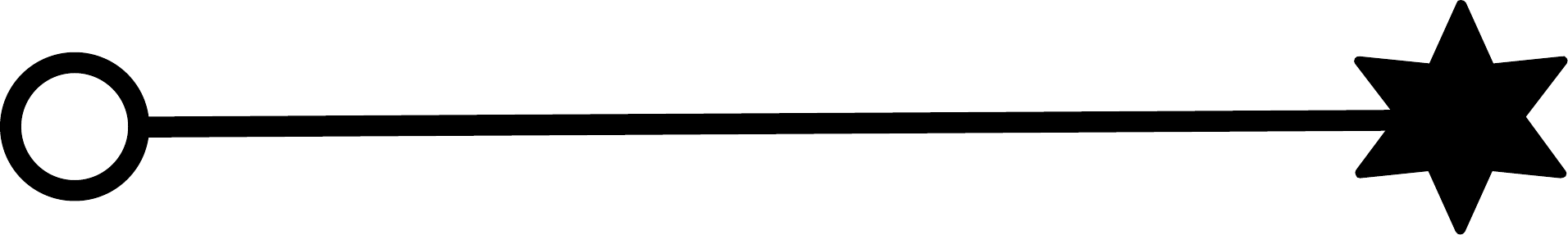}}{_q^\theta}}_,
	\label{eq:Baez_pi_theta}
\end{align}
where we depicted the annihilation operator by an empty circle. The operator (\ref{eq:Baez_pi_theta}) now annihilates a particle at the point $p$, creates a holonomy along $e$ and creates a particle at the endpoint $q$ of the holonomy. Since the particles $\theta$ are indistinguishable, this operator effectively transports a particle lying at $p$ along $e$ to $q$. 

For the sake of completion, we also show the last variant of the path observables including two spin $\frac{1}{2}$ fermions and one holonomy
\begin{align}
	\hat{\theta}^\dagger_{p \, A} \tensor{\jrep{\frac{1}{2}}{h_e}}{^A_B} \epsilon^{BC} \hat{\theta}_{q C}^{\dagger} = \,\, \overset{j = \frac{1}{2}}{\leftidx{_p^{\theta^\dagger}}{\includegraphics[valign=c, scale=0.12]{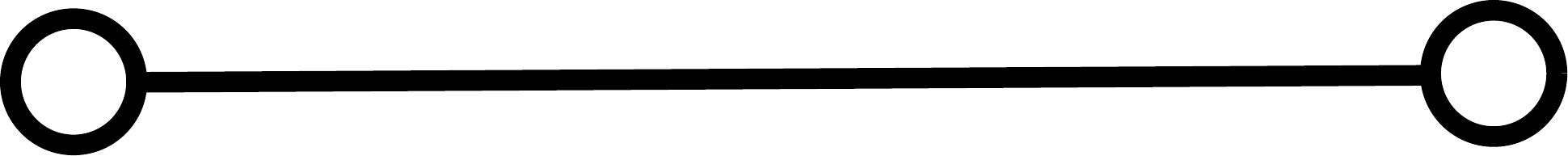}}{_q^{\theta^\dagger}}}_.
	\label{eq:Baez_pi_pi}
\end{align}
Given a particle of higher spin, we can define the creation two matter fields, for instance
\begin{align}
	\hat{\theta}^{\mathcal{A}}_p \epsilon_{\mathcal{A} \mathcal{B}} \tensor{\jrep{j}{h_e}}{^{\mathcal{B}}_{\mathcal{C}}} \hat{\theta}^{\mathcal{C}}_q,
\end{align}
which creates two spin $j$ particles at the points $p, q \in \Sigma$ and a spin $j$ holonomy inbetween.

The last operator we want to draw attention to can be obtained as a special case of (\ref{eq:Baez_pi_theta}) by choosing a trivial edge and $p = q$ and consequently also a trivial holonomy $\jrep{\frac{1}{2}}{h_e} = \mathds{1}$. We end up with the operator
\begin{align}
	\hat{\theta}^\dagger_{p \, A} \hat{\theta}^A_p = \,\, \leftidx{^{\theta^\dagger}}{\includegraphics[valign=c, scale=0.1]{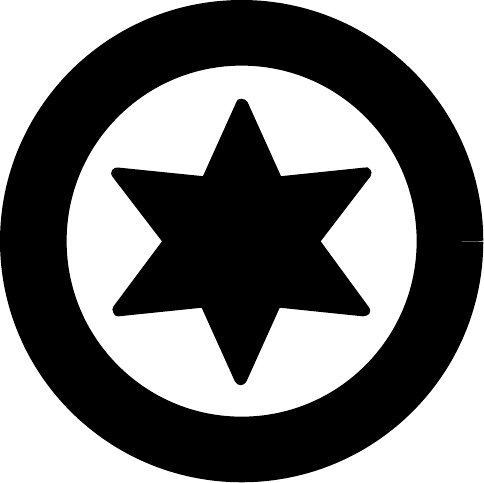}}{_p^\theta} =: \hat{N} + \mathds{1},
	\label{eq:Baez_number}
\end{align}
which creates and afterwards annihilates a particle $\theta$ at the point $p$. This operator is related to the well known number operator $\hat{N}$ in quantum field theory. It can be also shown \cite{Krasnov:1995vn} that the (anti)commutator of the path observables (\ref{eq:Baez_theta_theta} - \ref{eq:Baez_pi_pi}) and (\ref{eq:Baez_number}) form again path observables if the holonomies meet at endpoints. This statement still holds when using the scalar creation and annihilation operators. For two edges $e(0) = p, e(1) = f(0) = q, f(1) = r$, for example, we find
\begin{widetext}
	\begin{align}
		\comm{\stackrel{e}{\leftidx{_p^\theta}{\includegraphics[valign=c, scale=0.12]{Singlet.pdf}}{_q^{\theta}\, }}}{\, \stackrel{f}{\leftidx{_q^{\theta^\dagger}}{\includegraphics[valign=c, scale=0.12]{Baez_pi_theta.pdf}}{_r^\theta}}}_\pm = \, \stackrel{e \, \circ \, f}{\leftidx{_p^\theta}{\includegraphics[valign=c, scale=0.12]{Singlet.pdf}}{_r^{\theta}\, }}_,
		\label{eq:Baez_comm}
	\end{align}
\end{widetext}
which is a direct consequence of the canonical (anti)commutation relations of $\theta, \theta^\dagger$ and $h_e$. One can read (\ref{eq:Baez_comm}) such that stars and open circles can be linked to a longer holonomy. Indeed \cite{Baez:1997bw, Krasnov:1995vn}, this rule applies to all possible combinations of path observables. If there are multiple possibilities to link stars and circles, we will be left with a linear combination of those links. On the other hand, if there is no such possibility, then the (anti)commutator vanishes. (Anti)Commutators involving the number operator also behave in the same way, for instance
\begin{align}
	\comm{\leftidx{_q^{\theta^\dagger}}{\includegraphics[valign=c, scale=0.12]{Baez_number.pdf}}{^\theta} \,}{\, \stackrel{f}{\leftidx{_q^{\theta^\dagger}}{\includegraphics[valign=c, scale=0.12]{Singlet.pdf}}{_r^\theta}}}_\pm = \, \stackrel{f}{\leftidx{_q^\theta}{\includegraphics[valign=c, scale=0.1]{Singlet.pdf}}{_r^{\theta}\, }}_.
	\label{eq:Baez_comm_number}
\end{align}
With the operators introduced in this section we have found an intuitive formulation of creation and annihilation of parts of generalized spin network states including matter fields. As a next step, we want to take a look at the diffeomorphism constraint and its effect on the quantum states.

\section{Diffeomorphism Symmetry}
\label{sec:diffeo}
On the way to a quantum theoretical formulation of matter fields, we need to understand the symmetries of the theory. The generators of these symmetries are besides the Gauss constraint -- on a kinematical level -- also the diffeomorphism constraint. In this section, we will discuss the action of the diffeomorphism constraint in the quantum theory, and the symmetry properties it imposes upon states. We refer to Appendix \ref{app:spatial_diffeo} for a review of the standard derivation of the diffeomorphism constraint for vacuum loop quantum gravity \cite{thiemann_2007, Ashtekar_1995}. There we extend it by the contribution from Dirac theory of spin $\frac{1}{2}$ particles \cite{Thiemann_1998_QSD, Bojowald_2008} and discuss details of the separation of gauge and diffeomorphism symmetry.

\subsection{Diffeomorphism Invariant Spin Network States with Matter Fields}
In generally covariant theories, the diffeomorphism symmetry ensures that physical information may only be extracted from the equivalence classes of diffeomorphism invariant states. In particular, the absolute point inside the spatial hypersurface $p \in \Sigma$ has no physical relevance, rather we can deform $\Sigma$ by semi-analytic diffeomorphisms and do not change any physical observable. In matter-free loop quantum gravity, states fulfilling the diffeomorphism constraint are obtained via an averaging method \cite{Ashtekar_1995, Giulini:1999kc, Ashtekar_2004} yielding equivalence classes of deformed states. 

It will make a significant difference what conditions are imposed on the diffeomorphisms which constitute the diffeomorphism symmetry. This has been discussed already in previous work \cite{Ashtekar_1995, Fairbairn:2004qe}. We will work in the semi-analytic category \cite{Lewandowski:2005jk}.

In the following, we will be particularly interested in the interplay between the (anti)symmetrization imposed on quantum states that are based on graphs with symmetries by the diffeomorphism constraint on the one hand, and the (anti)symmetrization of the state due to the statistics of the matter field on the other hand. The fact that (anti)symmetry is imposed in some cases due to the diffeomorphism constraint is a novel feature in loop quantum gravity. 

The spin network decomposition of the Hilbert space $\mathcal{H}_G$ contains all the spin network states that admit matter fields at the vertices of $\gamma$ such that they are lying in the kernel of the Gauss constraint. Let $\Diff{(\Sigma)}$ denote the group of semi-analytic diffeomorphisms. We will consider the following subgroups
\begin{align}
	\Diff_\gamma &= \{ \phi \in \Diff(\Sigma) \big| \phi(\gamma) = \gamma \} \\
	\TDiff_\gamma &= \Big\{ \phi \in \Diff_\gamma \big| \phi(e) = e \text{ and } \phi(v) = v \nonumber \\
	&\qquad \forall e \in E(\gamma), v \in V(\gamma) \Big\} \\
	\GS_\gamma &= \faktor{\Diff_\gamma}{\TDiff_\gamma}.
\end{align}
The group $\Diff_\gamma$ consists of those diffeomorphisms which map the graph $\gamma$ onto itself, while $\TDiff_\gamma$ maps the graph trivially onto itself, i.e. it maps every edge $e \in E(\gamma)$ and every vertex $v \in V(\gamma)$ onto itself. Note that this also ensures that all fermions are mapped along with the vertex they are attached to in the first place, as the diffeomorphism constraint can be reduced to spatial diffeomorphisms. The quotient of $\Diff_\gamma$ and $\TDiff_\gamma$ again forms a group whose elements we call graph symmetries $\phi \in \GS_\gamma$. These describe the permutations of edges and vertices within the graph $\gamma$, which can be achieved by a semi-analytic diffeomorphism. In particular, the number of graph symmetries $\#\GS_\gamma$ is finite.

We define the diffeomorphism invariant states by averaging over all the diffeomorphisms $\phi \in \Diff(\Sigma)$ taking two steps. First we define the action of a projection operator $\hat{P}_\gamma$ on a spin network state $\Psi_\gamma$ corresponding to the graph $\gamma$
\begin{align}
	\hat{P}_\gamma \Psi_\gamma := \frac{1}{\#\GS_\gamma} \sum_{\phi \in \GS_\gamma} \hat{U}_\phi \Psi_\gamma.
\end{align}
Here, $\hat{U}_\phi$ acts on the spin network state by mapping edges of holonomies and vertices of intertwiner as well as the vertices where a matter field is attached:
\begin{align}
	\hat{U}_\phi h_e = h_{\phi(e)} \qquad \hat{U}_\phi \iota = \iota \qquad \hat{U}_\phi \theta_p = \theta_{\phi(p)}.
\end{align}
In the second step, we average over the rest of the diffeomorphism group, namely the diffeomorphisms $\phi \in \faktor{\Diff{(\Sigma)}}{\Diff_\gamma}$, which move the graph $\gamma$. This group, however, has infinite cardinality, such that we have to define the state $\gbra{\eta(\Psi_\gamma)}$ in the dual space $\mathcal{H}^*_G$ with the following action on a spin network state $\Phi_\alpha \in \mathcal{H}$
\begin{align}
	\gscpr{\eta(\Psi_\gamma)}{\Phi_\alpha} = \sum_{\phi \in \T_\gamma} \scpr{\hat{U}_\phi \hat{P}_\gamma \Psi_\gamma}{\Phi_\alpha}. \label{eq:dual_diff_inv}
\end{align}
In the literature, $\eta$ is called the \textit{rigging map} \cite{Ashtekar_1995, Ashtekar_2004}. Note that as $\Phi_\alpha$ has a convergent norm, also the sum in (\ref{eq:dual_diff_inv}) will be convergent such that $\gbra{\eta(\Psi_\gamma)}$ is well-defined.\\[2 ex]
We want to understand the behavior of spin network states under diffeomorphism symmetry. We want to discuss a condition under which a spin network state will be mapped to 0. This way, it is possible to identify states that do not appear in nature as they are annihilated by diffeomorphism symmetry. We want to discuss one sufficient condition such that given a spin network state $\Psi_\gamma$ the group averaged state $\gbra{\eta(\Psi_\gamma)}$ vanishes\footnote{This was pointed out to one of the authors by Lewandowski \cite{Lewandowski_private} in the context of loop quantum gravity without matter.}.
\begin{lemma}
	\label{lem:diff_kill}
	Let $\gamma$ be a spin network graph and $\Psi_\gamma$ a spin network state. If there exists a graph symmetry $\psi \in \GS_\gamma$ such that $\hat{U}_{\psi} \Psi_\gamma = - \Psi_\gamma$. Then the diffeomorphism averaged state vanishes, in other words
	\begin{align}
		\gbra{\eta(\Psi_\gamma)} = 0.
	\end{align}
	\begin{proof}
		We prove the hypothesis by proving that the projection $\hat{P}_\gamma \Psi_\gamma$ vanishes. As $\GS_\gamma$ is a group with finitely many elements, we can rearrange the averaging sum in the following way
		\begin{align}
			\hat{P}\gamma \Psi_\gamma &= \frac{1}{\#\GS_\gamma} \sum_{\phi \in \GS_\gamma} \hat{U}_\phi \Psi_\gamma \nonumber \\
			&= \frac{1}{2 \#\GS_\gamma} \left( \sum_{\phi \in \GS_\gamma} \hat{U}_\phi \Psi_\gamma + \sum_{\phi \in \GS_\gamma} \hat{U}_{\phi \circ \psi} \Psi_\gamma \right) \nonumber \\
			&= \frac{1}{2 \#\GS_\gamma} \left( \sum_{\phi \in \GS_\gamma} \hat{U}_\phi \Psi_\gamma + \sum_{\phi \in \GS_\gamma} \hat{U}_{\phi} \hat{U}_{\psi} \Psi_\gamma \right) \nonumber \\
			&= \frac{1}{2 \#\GS_\gamma} \left( \sum_{\phi \in \GS_\gamma} \hat{U}_\phi \Psi_\gamma - \sum_{\phi \in \GS_\gamma} \hat{U}_{\phi} \Psi_\gamma \right) = 0,
		\end{align}
		where we used the group homomorphism property $\hat{U}_{\psi \circ \phi} = \hat{U}_\psi \hat{U}_\phi$, which becomes apparent from the definition of $\hat{U}$. In the first step, we  permuted the finitely many addends of the second sum adequately. It follows that also $\gbra{\eta(\Psi_\gamma)} = 0$.
	\end{proof}
\end{lemma}
In order to identify states which get annihilated due to the diffeomorphism symmetry, it suffices to find one graph symmetry which maps the spin network state $\Psi_\gamma$ onto its own negative. As we require physical states to be invariant under diffeomorphism symmetry, we may call the states which satisfy the condition of Lemma \ref{lem:diff_kill} unphysical. Note, however, that this condition might not be necessary for having an unphysical state, as we can also imagine multiple addends canceling only in the ensemble but not two terms alone. In the following, we will focus on the condition characterized by Lemma \ref{lem:diff_kill}.

\subsection{A Specific Spin Network Graph}
\label{subsec:specific_SNS}
If we consider the spin-statistics theorem known from quantum field theory on curved spacetime \cite{Verch:2001bv} as a guiding principle, we can study the behavior of spin network states under exchange of fermions or bosons by permuting the respective vertices via a graph symmetry. The exchange of the particles as well as the permutation of edges on the graph will yield signs which may lead to the condition needed for Lemma \ref{lem:diff_kill}. As spin network graphs can, in general, be very asymmetric, i.e. there might be only few or no nontrivial graph symmetries, we can hardly make statements about the physicality of general spin network states. Because of this, we start with the simplest spin network state admitting two fermions. In Appendix \ref{app:General_SNS}, we consider a more general class of spin network states. Let us consider a state which consists of an edge $e$ with a vertex at the starting- and endpoint each
\begin{align}
	\Psi_e &= \hat{\theta}^A_p \epsilon_{AB} \tensor{\jrep{\frac{1}{2}}{h_e}}{^B_C} \hat{\theta}^C_q \vac \nonumber \\
	&= C \cdot \stackrel{j = \frac{1}{2}}{\leftidx{_p}{\includegraphics[valign=c, scale=0.12]{Singlet.pdf}}{_q}}.
	\label{eq:two_spin_1/2}
\end{align}
Here, $\theta^A$ creates a spin $\frac{1}{2}$ fermion in the Weyl representation. The state $\Psi_e$ contains two fermions at two distinct points $p, q \in \Sigma$. Up to a constant $C$, which is determined by normalization of the intertwiner, we can depict the algebraic formula of the spin network state by a spin network graph. The fermions sitting at the intertwiners at $p$ and $q$ are depicted by a star. Although, this graphical notation is very similar to the notation in \cite{Baez:1997bw}, there is a subtle difference between (\ref{eq:two_spin_1/2}), which is a state, and (\ref{eq:Baez_theta_theta}), which is an operator, having the same graphical representation. When being applied to the vacuum state\footnote{The vacuum state is the Ashtekar-Lewandowski vacuum in the gravitational sector and the Fock vacuum in the matter sector. The former is uniquely fixed by spatial diffeomorphism invariance, see \cite{Lewandowski:2005jk, Fleischhack:2004jc}.}, the operator (\ref{eq:Baez_theta_theta}) yields the state (\ref{eq:two_spin_1/2}). Thus, the similarity in the notation is justified. 

The state (\ref{eq:two_spin_1/2}) can be generalized to a system of two spin $j \in \frac{\mathds{N}_0}{2}$ particles with a suitable gravitational interaction inbetween
\begin{align}
	\Psi_e &= \theta^{\mathcal{A}}_p (\iota_{p})_{{\mathcal{A}} {\mathcal{B}}} \tensor{\jrep{j}{h_e}}{^{\mathcal{B}}_{\mathcal{C}}} \theta^{\mathcal{C}}_q \nonumber \\
	&= C \cdot \stackrel{j}{\leftidx{_p}{\includegraphics[valign=c, scale=0.12]{Singlet.pdf}}{_q}},
	\label{eq:two_spin_1/2_general}
\end{align}
where now $\tensor{\jrep{j}{h_e}}{^{\mathcal{A}}_{\mathcal{B}}}$ is a spin $j$ representation of SU(2), $\iota: j \otimes j \to 0$ is the intertwining operator and $\theta$ is a vector in the spin $j$ representation space of SU(2). The state $\Psi_e$ has one graph symmetry $\phi: \Sigma \to \Sigma$, which maps
\begin{align}
	\phi: p \mapsto q \qquad \phi: q &\mapsto p \qquad \phi: e \mapsto e^{-1}.
\end{align}
The diffeomorphism $\phi$ exchanges the two particles and hence is a good candidate for fulfilling the conditions of Lemma \ref{lem:diff_kill}. Also here, the state (\ref{eq:two_spin_1/2_general}) can be read in the two different ways, as a creation operator or as the state per se. In the following, we will think about it as creation operators, albeit it will not make a difference taking the opposite perspective. If $\hat{\Psi}$ creates the state $\Psi$, then the action of the Rigging map $\eta$ on $\hat{\Psi}$ can be expressed via the action on $\Psi$ in the following way
\begin{align}
	\eta\left( \hat{\Psi} \ket{0} \right) = \eta\left( \ket{\Psi} \right),
\end{align}
where $\ket{0}$ denotes the vacuum state. Let us apply the diffeomorphism $\phi$ to (\ref{eq:two_spin_1/2_general}) and get
\begin{align}
	\hat{U}_\phi \Psi_e &= \theta^{\mathcal{A}}_{\phi(p)} (\iota_{\phi(p)})_{{\mathcal{A}} {\mathcal{B}}} \tensor{\jrep{j}{h_{\phi(e)}}}{^{\mathcal{B}}_{\mathcal{C}}} \theta^{\mathcal{C}}_{\phi(q)} \nonumber \\
	&= \theta^{\mathcal{A}}_q (\iota_{q})_{{\mathcal{A}} {\mathcal{B}}} \tensor{\jrep{j}{h_e^{-1}}}{^{\mathcal{B}}_{\mathcal{C}}} \theta^{\mathcal{C}}_p.
	\label{eq:diff_two_spin}
\end{align}
In order to convert this expression into something comparable to (\ref{eq:two_spin_1/2_general}), we have to better understand the intertwiner $\iota_{\mathcal{A} \mathcal{B}}$ which couples two general spin $j$ holonomies. To do this, we can use the fundamental representation of SU(2) to build up any other irreducible representation. In particular, we find the following
\begin{lemma}
	Let $\iota: j \otimes j \to 0, \, j \in \frac{\mathds{N}_0}{2}$ be a gauge invariant intertwiner of SU(2). It holds
	\begin{align}
		\iota_{{\mathcal{A}} {\mathcal{B}}} = (-1)^{2j} \iota_{{\mathcal{B}} {\mathcal{A}}},
	\end{align}
	i.e. the intertwiner is symmetric for integral spin and anti-symmetric for half-integral spin.
	\begin{proof}
		We will proof this by giving an explicit construction of $\iota$. At first, note that the subspace of gauge invariant intertwiner, which couple
		\begin{align}
			j\otimes j \cong 0 \oplus 1 \oplus ... \oplus 2j
		\end{align}
		is one-dimensional. If we therefore find one gauge-invariant intertwiner $\iota$ as described above, it will be unique up to normalization.
		
		We rewrite the indices $\mu \nu$ as a number of $2j$ symmetrized spin $\frac{1}{2}$ indices and make the educated guess
		\begin{align}
			\iota_{(A_1 ... A_{2j})(B_1 ... B_{2j})} = \epsilon_{(A_1|(B_1|} ... \epsilon_{|A_{2j})|B_{2j})}, \label{eq:spin_jj_intertwiner}
		\end{align}
		where we denote the symmetrization grouping by a vertical line $|$. Hence, (\ref{eq:spin_jj_intertwiner}) is symmetric in $A_1, ..., A_{2j}$ as well as in $B_1, ..., B_{2j}$. To prove the intertwining property, we will successively use the intertwining property of $\epsilon$
		\begin{align}
			\tensor{\jrep{\frac{1}{2}}{g}}{^A_C} \tensor{\jrep{\frac{1}{2}}{g}}{^B_D} \epsilon_{AB} = \epsilon_{CD},
		\end{align}
		with an arbitrary element $g \in $ SU(2). If we now build $\jrep{j}{g}$ from $\jrep{\frac{1}{2}}{g}$ analogously, we end up with the desired intertwining property
		\begin{align}
			&\iota_{(A_1 ... A_{2j})(B_1 ... B_{2j})} \tensor{\jrep{\frac{1}{2}}{g}}{^{A_1}_{C_1}} \tensor{\jrep{\frac{1}{2}}{g}}{^{B_1}_{D_1}} \cdots \times \nonumber \\
			&\times \cdots \tensor{\jrep{\frac{1}{2}}{g}}{^{A_{2j}}_{C_{2j}}} \tensor{\jrep{\frac{1}{2}}{g}}{^{B_{2j}}_{D_{2j}}} = \iota_{(C_1 ... C_{2j})(D_1 ... D_{2j})} \nonumber \\
			&\iff \iota_{{\mathcal{A}} {\mathcal{B}}} \tensor{\jrep{j}{g}}{^{\mathcal{A}}_{\mathcal{C}}} \tensor{\jrep{j}{g}}{^{\mathcal{B}}_{\mathcal{D}}} = \iota_{{\mathcal{C}}{\mathcal{D}}},
		\end{align}
		which proves (\ref{eq:spin_jj_intertwiner}). Finally, we can read off
		\begin{align}
			\iota_{(A_1 ... A_{2j})(B_1 ... B_{2j})} &= (-1)^{2j} \iota_{(B_1 ... B_{2j})(A_1 ... A_{2j})} \nonumber \\
			\iota_{{\mathcal{A}} {\mathcal{B}}} &= (-1)^{2j} \iota_{{\mathcal{B}} {\mathcal{A}}},
		\end{align}
		yielding a sign factor $-1$ for each of the $\epsilon$ in (\ref{eq:spin_jj_intertwiner}).
	\end{proof}
	\label{lem:spin_jj_intertwiner}
\end{lemma}
We are now ready to take a closer look at (\ref{eq:diff_two_spin}) and compare it to (\ref{eq:two_spin_1/2_general})
\begin{align}
	\hat{U}_\phi \Psi_e &= (\iota_{q})_{{\mathcal{A}} {\mathcal{B}}} \tensor{\jrep{j}{h_e^{-1}}}{^{\mathcal{B}}_{\mathcal{C}}} \left( \theta^{\mathcal{A}}_q \theta^{\mathcal{C}}_p \right) \nonumber \\
	&= \tensor{\jrep{j}{h_e}}{^{\mathcal{B}}_{\mathcal{A}}} (\iota_{q})_{{\mathcal{B}} {\mathcal{C}}} \left( (-1)^{2j}  \theta^{\mathcal{C}}_p \theta^{\mathcal{A}}_q \right) \nonumber \\
	&= (-1)^{4j} \tensor{\jrep{j}{h_e}}{^{\mathcal{B}}_{\mathcal{A}}} (\iota_{q})_{{\mathcal{C}} {\mathcal{B}}}  \theta^{\mathcal{C}}_p \theta^{\mathcal{A}}_q \nonumber \\
	&= \Psi_e,
	\label{eq:diff_action_two_spins}
\end{align}
where we used the (anti)commutation relations of $\theta$ in the first step, the intertwining property in the second step, and Lemma \ref{lem:spin_jj_intertwiner} in the last step. From (\ref{eq:diff_action_two_spins}) we deduce that the simple spin network graph survives the diffeomorphism group averaging. However, if we were to choose the opposite statistics, the state would lie inside the kernel of the rigging map $\eta$. Note that the above considerations also hold true for any disjoint union of an arbitrary spin network state with a pair of particles (\ref{eq:two_spin_1/2_general}).

\subsection{Behavior of General Spin Network States}
It quickly becomes apparent that we cannot ensure that there always exists a diffeomorphism which performs the desired exchange of particles. A counter example can be constructed from (\ref{eq:two_spin_1/2_general}) just by gauge invariantly coupling a spin network on one of the star vertices but not on the other one. This state can be written as
\begin{align}
	\Psi_\text{asymm} &= F^{\mathcal{D}} \theta^{\mathcal{A}}_p (\iota_{p})_{{\mathcal{D}} {\mathcal{A}} {\mathcal{B}}} \tensor{\jrep{j}{h_e}}{^{\mathcal{B}}_{\mathcal{C}}} \theta^{\mathcal{C}}_q \nonumber \\
	&= \stackrel{\qquad \qquad j}{\leftidx{_{j\pm \frac{1}{2}}}{\includegraphics[valign=c, scale=0.12]{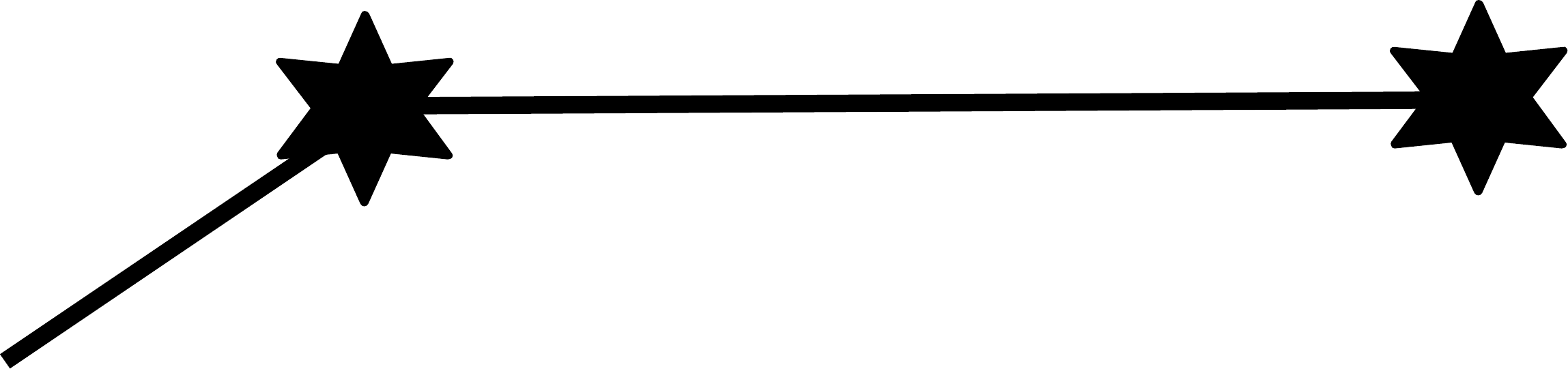}}{_,}}
\end{align}
where $F^{\mathcal{D}}$ denotes a spin network graph which couples to the rest gauge invariantly. As we can see, there is no way to exchange the two particles without changing the topology of the graph. Still, we can study the behavior on more general spin network graphs, which do admit a graph symmetry exchanging two particles. In Appendix \ref{app:General_SNS}, we challenge this idea and construct the most general spin network states admitting two particles and a graph symmetry which exchanges the two. 

There, we found rules for the detailed construction of states which survive the diffeomorphism constraint. Given a generic spin network state with graph symmetry, we can decompose the intertwiner on the symmetry axis into the Wigner basis and deduce from the statistics of the particles which components will survive or vanish after group averaging. Specifically, fermions will let those components survive which gather a total minus sign and bosons will let those components survive which are invariant under the diffeomorphism $\phi_\text{general}$.

\section{Classical Actions}
\label{sec:actions}
As we have seen, quantum matter fields can be naturally coupled to holonomies which represent the fundamental building blocks of quantum geometry and are built from representations of SU(2). To embed the discussions of the previous sections into a more complete picture, we want to discuss the corresponding classical theory from where we would start a canonical quantization in the first place.

For the classical theory of spin $\frac{1}{2}$ fields, we will review the work \cite{Thiemann_1998_QSD, Thiemann_1998, Bojowald_2008}. Subsequently, we generalize the idea to spin 0 as well as higher spin fields. For this, we follow the same steps of the canonical quantization program as in the vacuum theory but now for an action 
\begin{align}
	S = S_\text{Gravity} + S_\text{Matter}.
\end{align}
For spin $\frac{1}{2}$ there is the well known Dirac action and the action for higher spin fields was already investigated early on \cite{Fiertz_39, Rarita_41}.

Note that the experimentally confirmed theories of integral spin particles are actually Yang-Mills theories of connection 1-forms. This puts us into a dilemma of choosing between a Fock quantization of integral spin creators and annihilators or a loop quantization of the holonomies defined by path ordered exponentials of connections (analogous to gravity degrees of freedom), which would be a natural choice within the background-independent loop quantization of gravity. While a Fock quantization would fit better into the particle picture of the previous sections, the loop quantization of a Yang-Mills action with an underlying gauge symmetry group $G$ is already well understood (see for instance \cite{Corichi:1997us} and \cite{Ashtekar_1992}). However, the corresponding $G$-holonomies decouple from the gravitational holonomies such that the spin interaction character of the particles is lost and the classical derivation would be inconsistent with the quantum theory discussed above. Note that not only on a quantum level the two perspectives yield different theories, but also the classical theories turn out to be inequivalent (see Appendix \ref{app:U1_action} for details).

\subsection{Dirac Spinors}
In a gravity theory of fermionic matter, we can achieve a Hamiltonian formulation. The Dirac action for a spin $\frac{1}{2}$ particle $\Psi$ and its conjugate momentum $\Pi := \sqrt{\det(g)} \Psi^\dagger$ reads
\begin{align}
	S_\text{Dirac} = \frac{i}{2} \int_\mathcal{M} d^4x \sqrt{- \det(g)} \left( \overline{\Psi} \gamma^\alpha e^\mu_\alpha \nabla_\mu \Psi - \overline{ \nabla_\mu \Psi} \gamma^\alpha e^\mu_\alpha \Psi \right),
	\label{eq:Dirac_act_original}
\end{align}
with $\overline{\Psi} = \Psi^\dagger \gamma^0$, $e^\mu_\alpha$ being the tetrad field and $\nabla$ the covariant derivative which annihilates $e$. The spatial part of the corresponding connection can later be identified with the Ashtekar connection $A$ \cite{Thiemann_1998_QSD}. If we now introduce a foliation described by lapse function $N$ and shift vector field $N^a$ via $n^\mu = \frac{1}{N} (T^\mu - N^\mu)$ and $n_\alpha = - \delta_{\alpha}^0$. We can then write the tetrad as
\begin{align}
	e^\mu_\alpha = i^*(e)^\mu_\alpha - n_\alpha n^\mu,
\end{align}
with the triad $e^a_i := i^*(e)^a_i$ being the pullback of the tetrad onto the spatial hypersurface $\Sigma$. We also decompose the Dirac spinor into two chiral components $\Psi = (\psi, \eta)$, which are both Weyl spinors. Plugging in all these quantities and splitting up the derivatives into the ones along the time vector field $T^\mu$ and the spatial derivatives $\mathcal{D}_a$, the Dirac action takes the form of a constrained system with Dirac contributions to the Gauss, diffeomorphism and Hamilton constraint \cite{Thiemann_1998_QSD}.

The canonical variables are now given by the chiral components $(\psi, \eta)$ and their conjugate momenta $(\pi_\psi, \pi_\eta) = i \sqrt{\det(g)}(\psi^\dagger, \eta^\dagger)$ satisfying the anti-Poisson relations
\begin{align}
	\left\{ \psi^{A} (x), \psi^{B} (y) \right\}_{\pm} = 0 \qquad &\left\{ \pi_{\psi \, A} (x), \pi_{\psi \, B} (y) \right\}_{\pm} = 0 \\
	\left\{ \psi^{A} (x),
	\pi_{\psi \, B} (y) \right\}_{\pm} &= \delta^{A}_{B} \delta_{x, y},
\end{align}
and similar for $\eta$. Also the anti-Poisson relations which mix $\psi$ and $\eta$ vanish. Note that the momenta $\pi_\psi$ go with a relative factor $\sqrt{\det(q)}$ in comparison to the matter fields $\psi$. It turns out that the discussion is being simplified by transforming the (spacetime scalar) Weyl spinors to half-densities \cite{Thiemann_1998}
\begin{align}
	\xi = \sqrt[4]{\det(q)} \psi \quad \text{ and } \rho = \sqrt[4]{\det(q)} \eta.
\end{align}
As a consequence, the conjugate momenta $\pi_\xi, \pi_\rho$ are also half-densities. They satisfy the simple reality conditions
\begin{align}
	\pi_\xi = i \xi^\dagger \qquad \pi_\rho = i \rho^\dagger.
	\label{eq:reality_condition}
\end{align}
With both contributions, Holst and Dirac, the constraints can be written in the following form \\
\begin{align}
	G_i &= \frac{1}{\kappa} \mathcal{D}_a E^a_i + i (\xi^\dagger \tau_i \xi + \rho^\dagger \tau_i \rho) \label{eq:Dirac_Gauss} \\
	V_a &= \frac{1}{\kappa} F_{ab}^i E^b_i + \frac{i}{2} \left( \xi^\dagger \mathcal{D}_a \xi + \rho^\dagger \mathcal{D}_a \rho - c.c. \right) \label{eq:Dirac_Diffeo} \\
	H &= \frac{1}{2\kappa \sqrt{q}} \left( (2 [K_a, K_b]^i - F_{ab}^i) [E^a, E^b]_i \right) + \nonumber \\
	&+ \frac{E^a_i}{2 \sqrt{\det(q)}} \Big( \mathcal{D}_a (\xi^\dagger \sigma^i \xi + \rho^\dagger \sigma^i \rho) + \nonumber \\
	&+ i (\xi^\dagger \sigma^i \mathcal{D}_a \xi - \rho^\dagger \sigma^i \mathcal{D}_a \rho - c.c.) + \nonumber \\
	&- K^i_a (\xi^\dagger \xi - \rho^\dagger \rho) \Big),
	\label{eq:Dirac_Hamilton}
\end{align}
where the constant $\kappa$ has to be taken into account since $S_\text{Holst}$ and $S_\text{Dirac}$ carry different units. 

The constraints can then be interpreted as the generators of gauge transformations as in matter-free loop quantum gravity. The kinematical Hilbert space can then be finally set up as a tensor product space of cylindrical functions of holonomies $h_e$ together with the antisymmetric Fock space $\mathcal{F}^- (\mathcal{h}^{(j)})$ as discussed in \autoref{sec:LQG_matter}. For a way to write the Fock space in which the states are (wave) functions, in keeping with the gravitational Hilbert space, see \cite{Thiemann_1998}.

\subsection{Integral Spin Quantum Fields}
\label{subsec:Integ}
\subsubsection{The Spin 0 Field}
Before we consider higher spin quantum fields, we want to shortly discuss the classical theory of a spin 0 and a spin 1 field explicitly. As a first step, let us take a look at the real, massless Klein-Gordon field described by the Klein-Gordon action
\begin{align}
	S_\text{KG} = - \frac{1}{2} \int_\mathcal{M} d^4x \sqrt{-\det(g)} g^{\mu \nu} \mathcal{D}_\mu \phi \mathcal{D}_\nu \phi.
	\label{eq:spin0_action}
\end{align}
Since the matter fields $\phi$ are spacetime and SU(2) scalars, the covariant derivatives can also be replaced by partial derivatives.

We can perform a Legendre transformation, going over to Arnowitt-Deser-Misner (ADM) variables \cite{Arnowitt:1962hi} for the metric and to $\phi$ and its conjugate momentum
\begin{align}
	\pi &= \frac{\partial \mathcal{L}}{\partial (\partial_0 \phi)}= \frac{\sqrt{\det(q)}}{N} \left( - \partial_0 \phi + N^a \partial_a \phi \right)
\end{align}
The term $\pi \dot{\phi}$ can be manipulated in such a way that the canonical variables are both half-densities 
\begin{align}
	\xi = \sqrt[4]{\det(q)} \phi \quad \text{ and } \quad \pi_\xi = \frac{\pi}{\sqrt[4]{\det(q)}}.
\end{align}
In this case, we will have to keep the covariant derivatives in order to absorb the half-density factor inside the matter field $\phi$. However, the transformation to the half-density $\xi$ changes the symplectic structure. We can go back to canonical variables by redefining the Ashtekar connection. 
In the symplectic structure, we gather the following excess term due to the product rule:
\begin{align}
	\pi \dot{\phi} = \pi_\xi \dot{\xi} - \pi_\xi \xi \frac{\mathcal{L}_t \left( \sqrt[4]{\det(q)} \right)}{\sqrt[4]{\det(q)}}.
\end{align}
The excess term can be reformulated in terms of the flux variable $\dot{E}^a_i$. In order to arrive there, we write
\begin{align}
	\frac{\mathcal{L}_t \left( \sqrt[4]{\det(q)} \right)}{\sqrt[4]{\det(q)}} = \frac{1}{4} q^{ab} \dot{q}_{ab} = - \frac{1}{2} \dot{e}^a_i e_a^i.
	\label{eq:excess_term}
\end{align}
We can use both expressions of (\ref{eq:excess_term}) in the following linear combination
\begin{align}
	\frac{\mathcal{L}_t \left( \sqrt[4]{\det(q)} \right)}{\sqrt[4]{\det(q)}} &= \left( \frac{3}{2} - \frac{1}{2} \right) \frac{\mathcal{L}_t \left( \sqrt[4]{\det(q)} \right)}{\sqrt[4]{\det(q)}} \nonumber \\
	&= \frac{3}{2} \cdot \frac{1}{4} q^{ab} \dot{q}_{ab} + \frac{1}{2} \cdot \frac{1}{2} \dot{e}^a_i e_a^i \nonumber \\
	&= \frac{e_a^i}{4 \sqrt{\det(q)}} \dot{E}^a_i.
	\label{eq:excess_term_2}
\end{align}
We can combine the term (\ref{eq:excess_term_2}) with the symplectic term $- \dot{E}^a_i A_a^i$ coming from the gravity action. Finally, the new canonical variable, which is conjugate to $E$, reads
\begin{align}
	A^i_a \mapsto \tilde{A}^i_a = A^i_a + \pi_\xi \xi \frac{e_a^i}{4 \sqrt{\det(q)}}.
	\label{eq:new_connection}
\end{align}
This also means that every appearance of the Ashtekar connection $A$ in the Holst contributions of the constraints has to be replaced by the new connection $\tilde{A}$ minus the excess term.

We can finally write down the constraints in terms of the new canonical variables
\begin{align}
	V_a &= \pi_\xi \mathcal{D}_a \xi \label{eq:KG_diffeo} \\
	H &= \frac{1}{2} \left( q^{ab} \mathcal{D}_a \xi \mathcal{D}_b \xi + \pi_\xi^2 \right).
\end{align}
In these equations, $\mathcal{D}_a$ is the covariant derivative using the Levi-Civita connection, which can be expressed in terms of $E^a_i$.
One can see that -- as expected for a spin 0 field -- there is no contribution to the Gauss constraint, hence the gauge transformation of $\phi$ is trivial. The diffeomorphism constraint takes a similar form as in the Dirac case and can be proven to generate spatial diffeomorphisms. From this point on, we can follow the same steps as in the Dirac case and describe the quantum theory by a tensor product of the vacuum spin network states with a Fock space of spin 0 particles. Finally, we end up with a rather boring theory, since there is no constraint on the entanglement of the spin 0 particles with the gravitational holonomies but rather the particles may be placed at any point in $\Sigma$, also displaced from the spin network graph $\gamma$. As the spin 0 particle is gauge invariant, we would not expect anything different. 

\subsubsection{The Spin 1 Field}
A more interesting theory emerges from an action of the spin 1 particle. For the reasons discussed in the beginning of this section, we will refrain from starting with a Yang-Mills theory, although most of the experimentally confirmed matter theories are Yang-Mills theories of connection 1-forms in classical field theory of integral spin matter fields. Rather we choose a minimal action for a spin 1 particle inspired by the Klein-Gordon action. A massive spin 1 particle can be described by a rank 1 tensor $\phi_I$, $I \in \{0, 1, 2, 3\}$ and the action \cite{Singh:1974qz}
\begin{align}
	S &= \int_\mathcal{M} d^4x \frac{N}{2} \sqrt{\det(q)} \Big[- g^{\mu \nu} \eta^{IJ} \nabla_\mu \phi_I \nabla_\nu \phi_J + \nonumber \\
	&- m^2 \phi_I \phi^I + e^\mu_I \nabla_\mu \phi^I e^\nu_J \nabla_\nu \phi^J \Big],
	\label{eq:spin1_action}
\end{align}
where we lower and raise internal indices $I, J, ...$ with the Minkowski metric $\eta$ and $\nabla$ is the covariant derivative with the spin connection $\tensor{\omega}{_\mu^I_J}$, which defines a parallel transport of the spin 1 representations and is compatible with the tetrad $e^I_\mu$. 

Note that it is important for the following analysis that we consider matter fields with internal structure $\phi_I$ rather than spacetime tensors $A_\mu$. We show in Appendix \ref{app:U1_action} that the two theories are inequivalent in the presence of gravity. 

The action of the spin 1 field (\ref{eq:spin1_action}) yields the equations of motion
\begin{align}
	g^{\mu \nu} \nabla_\mu \nabla_\nu \phi_I - m^2 \phi_I - e^\mu_I e^\nu_J \nabla_\mu \nabla_\nu \phi^J = 0.
	\label{eq:spin_1_eom_blank}
\end{align}
If we contract (\ref{eq:spin_1_eom_blank}) with $\nabla_\rho e^\rho_K \eta^{KI} =: \nabla^I$, then we get the so called Lorentz condition
\begin{align}
	m^2 e^\mu_I \eta^{IJ} \nabla_\mu \phi_J =: m^2 \nabla^J \phi_J = 0.
	\label{eq:spin_1_Lorentz}
\end{align}
The Lorentz condition reduces the degrees of freedom by 1, such that we end up with 3 degrees of freedom as expected for a spin 1 particle. Note, however, that this formulation is only valid if we impose $m^2 \neq 0$. If we insert the Lorentz condition (\ref{eq:spin_1_Lorentz}) into the equations of motion (\ref{eq:spin_1_eom_blank}), we can derive the spin 1 equivalent of the Klein Gordon equation
\begin{align}
	g^{\mu \nu} \nabla_\mu \nabla_\nu \phi_I - m^2 \phi_I = 0.
	\label{eq:spin1_KG}
\end{align}
We can perform a Legendre transformation analogous to the spin 0 case before. The action can be written again in terms of three constraints 
\begin{align}
	S &= \int_\mathcal{M} d^4 x \, \pi^I \dot{\phi}_I - \Bigg[ \tensor{\omega}{_0^J_I} \pi^I \phi_J + N^a \pi^I \nabla_a \phi_I + \nonumber \\
	&- \frac{N}{2} \sqrt{\det(q)} \Bigg( q^{ab} \eta^{IJ} \nabla_a \phi_I \nabla_b \phi_J + \frac{\pi^I \pi_I}{\det(q)} + \nonumber \\
	&+ m^2 \phi_I \phi^I - e^a_i \nabla_a \phi^i \nabla^I \phi_I - \pi^0 \nabla^I \phi_I \Bigg) \Bigg],
	\label{eq:spin1_legendre}
\end{align}
where we again used the metric in terms of the shift vector field $N^a$ and the lapse function $N$ being the Lagrange multipliers for the diffeomorphism and Hamilton constraint and denoted the conjugate momentum to $\phi_I$ by $\pi^I$. As in the spin 0 case, we can then rearrange the half-density weights $\sqrt[4]{\det(q)}$ to define the new half-density matter field $\xi := \sqrt[4]{\det(q)} \phi$ and its conjugate momentum $\pi_\xi = \frac{\pi}{\sqrt[4]{\det(q)}}$ by extending the Ashtekar connection by an adequate excess term analogue to (\ref{eq:new_connection}) to make the symplectic structure invariant under this transformation. If we take a closer look at the variables $\phi_I$ and $\pi^I$, it turns out that one of the four degrees of freedom is fixed by a constraint. These couple with the constraints arising for the spin connection $\omega$.

In Appendix \ref{app:constraints}, we list these technical hurdles in detail. Although a satisfying solution to these problems is not known yet, reminiscent structures from previously discussed theories appear. This lets us conjecture the spin 1 field contributions to the Gauss, diffeomorphism and Hamilton constraint to take a similar form compared to the previous matter theories, namely
\begin{align}
	\tensor{G}{_i} &= \tensor{\epsilon}{_{ij}^k} \pi_\xi^j \xi_k \label{eq:spin1_Gauss} \\
	V_a &= \pi_\xi^i \mathcal{D}_a \xi_i \\
	H &= \frac{1}{2} \left( q^{ab} \mathcal{D}_a \xi_i \mathcal{D}_b \xi^i + \pi_\xi^i \pi_{\xi \, i} + m^2 \xi_i \xi^i \right),
	\label{eq:spin1_Hamilton}
\end{align}
where the covariant derivative $\mathcal{D}$ contains the Ashtekar connection $A$ and is compatible with the triad $e$. In order to formulate the constraints in terms of the extended Ashtekar connection $\tilde{A}$, one would have to choose another covariant derivative and collect correction terms. Note that we already left out the terms in the Hamilton constraint arising from the additional term $(\nabla \phi)^2$ of the action (\ref{eq:spin1_action}). The Gauss constraint $\tensor{G}{_i}$ can now be interpreted as the generator of gauge transformations of the spin 1 particle $\xi$. To see this, we calculate the Poisson bracket of the canonical variables $(\pi_\xi, \xi)$ with the smeared Gauss constraint $G(\Lambda)$
\begin{align}
	\{G(\Lambda), \phi_i\} &= - \tensor{\Lambda}{^j} \tensor{\epsilon}{_{ij}^k} \phi_k \\
	\{G(\Lambda), \pi^i\} &= \tensor{\Lambda}{^k} \tensor{\epsilon}{_{kj}^i} \pi^j.
\end{align}
The vector constraint $V_a$ and Hamilton constraint $H$ on the other hand, take a similar form as in the previous theories.

\subsubsection{Higher Integral Spin Fields}
Although there is still work to be done in the case of spin 1, this gives rise to the conjecture that similar matter theories with higher spin can be treated in a similar way. However, in general the introduction of auxiliary fields is needed to derive a generalized Klein-Gordon equation and the Lorentz conditions
\begin{align}
	g^{\mu \nu} \nabla_\mu \nabla_\nu \phi_{I_1 ... I_j} &= 0 \label{eq:general_Klein_Gordon} \\
	\eta^{I_1 J} e^\mu_J \nabla_\mu \phi_{I_1 ... I_j} &= 0,
	\label{eq:integral_Lorentz}
\end{align}
with $\phi$ being a symmetric tensor field, which describe an integral spin $j$ field when the Lorentz conditions (\ref{eq:integral_Lorentz}) are imposed. The auxiliary fields vanish on shell as long as we do not add potential terms. In addition to that, the minimal Lagrangian for an arbitrary integral spin $j$ field contains the terms (\ref{eq:spin1_action}) in addition to terms which are proportional to the auxiliary fields. According to \cite{Singh:1974qz} the Lagrangian for particles with integral spin reads
	\begin{align}
		\mathcal{L} &= \sqrt{-\det(g)} \frac{1}{2} \phi^{(j)}_{I_1 ... I_j} \left( g^{\mu \nu} \nabla_\mu \nabla_\nu - m^2 \right) \phi^{(j) \, I_1 ... I_j} + \nonumber \\
		&+ \sqrt{-\det(g)} \frac{j}{2} (\nabla \phi^{(j)})_{I_2 ... I_j} (\nabla \phi^{(j)})^{I_2 ... I_j} + \nonumber \\
		&+ \mathcal{O}(\phi^{(j - 1)})
		\label{eq:general_boson_action}
	\end{align}
where we pull internal indices with the Minkowski metric $\eta$ and defined $(\nabla \phi^{(j)})_{I_1 ... I_{j-1}} = \eta^{IJ} e^\mu_I \nabla_\mu \phi^{(j)}_{J I_1 ... I_{j-1}}$. The Landau symbol $\mathcal{O}(\phi^{(j-1)})$ collects all the terms which are at least linear in the introduced auxiliary fields $\phi^{(j-q)}$ with $j \geq q > 0$. These auxiliary fields vanish on shell. Moreover, the additional term $(\nabla \phi^{(j)})^2$ reduces the number of degrees of freedom of the matter field $\phi^{(j)}$ to $2j +1$. Note that this method only works for massive fields with $m^2 \neq 0$. We see from the first term of (\ref{eq:general_boson_action}) that we recover the same terms as before in the Gauss, diffeomorphism and Hamilton constraint when doing a constraint analysis. The remaining terms all contain at least one auxiliary field. Since we can reformulate the classical equations of motion in a way that the auxiliary fields vanish
\begin{align}
	\phi^{(j-q)} = 0 \qquad \forall j \geq q > 0,
	\label{eq:aux_vanish}
\end{align}
we will end up with the exact same form of the constraints (\ref{eq:spin1_Gauss} - \ref{eq:spin1_Hamilton}) after we implemented the conditions (\ref{eq:aux_vanish}). We conjecture that the most general form of the constraints will read
\begin{align}
	G_i &= \pi_\xi^{I_1 ... I_j} \tensor{\tau}{_i^{J_1 ... J_j}_{I_1 ... I_j}} \xi_{J_1 ... J_j} + \mathcal{O}(\phi^{(j-q)})
	\label{eq:general_Gauss} \\
	V_a &= \pi_\xi^{I_1 ... I_j} \mathcal{D}_a \xi_{I_1 ... I_j} + \mathcal{O}(\phi^{(j-q)}) \label{eq:general_diffeo} \\
	H &= q^{ab} \mathcal{D}_a \xi_{I_1 ... I_j} \mathcal{D}_b \xi^{I_1 ... I_j} + \pi_{\xi \, {I_1 ... I_j}} \pi_\xi^{I_1 ... I_j} + \nonumber \\
	&+ m^2 \xi_{I_1 ... I_j} \xi^{I_1 ... I_j} - e^a_i \mathcal{D}_a \xi^{i I_2 ... I_j} \mathcal{D}^I \xi_{I I_2 ... I_j} + \nonumber \\
	&- \pi_\xi^{0 I_2 ... I_j} \mathcal{D}^I \xi_{I I_2 ... I_j} + \mathcal{O}(\phi^{(j-q)}),
	\label{eq:general_Hamilton}
\end{align}
where we denoted any function of the auxiliary fields $\phi^{j-q}$ which vanishes when implementing the on shell conditions (\ref{eq:aux_vanish}) by $\mathcal{O}(\phi^{(j-q)})$ and $\tau_i$ being the basis elements of the Lie algebra $\mathfrak{su}(2)$ in the spin $j$ representation.

Note that the order of implementing the constraints becomes important now. Since we would like to implement the Gauss constraint as the generator of SU(2) gauge transformations, we can simplify this discussion by first projecting onto the subspace where (\ref{eq:aux_vanish}) holds. Otherwise, we would have to deal with the definition of a Fock space, in particular creation and annihilation operators for each of the auxiliary fields $\hat{\phi}^{(j-q)}$ and the action of the Gauss constraint would be more complicated.

The covariant phase space formalism \cite{Lee:1990nz} seems particularly useful for this endeavor, as it clearly brings out the points in the construction of the phase space in which the equations of motion can be used.

\subsection{Half-Integral Spin Quantum Fields}
We finally suggest a general theory of loop quantum gravity with particles of spin $j \in \frac{2\mathds{N}_0 + 1}{2}$. We are looking for a theory of gravity minimally interacting with matter fields, i.e. we want to find an action which yields the same equations of motion as in the free theory with covariant derivatives. Even on classical level, the interacting theories of higher spin particles is a current topic of interest \cite{Fronsdal:1978rb, Fang:1978wz, Vasiliev:2004qz, Bekaert:2010hw}. There are several unsolved problems including No-Go theorems for certain theories of higher spin particles. The very idea of a free theory of a half-integral spin particle can be traced back to \cite{Rarita_41}, which is in turn based on \cite{Fiertz_39}. The systematic approach based on this work needs the introduction of auxiliary matter fields, which vanish on shell in the free theory but not necessarily in an interacting theory.

There, a half-integer spin $n + \frac{1}{2} \in \mathds{N}_0 + \frac{1}{2}$ particle is described by a $\left(\frac{1}{2} (n+1), \frac{1}{2} n\right) \oplus \left( \frac{1}{2} n, \frac{1}{2} (n+1) \right)$ representation of the Lorentz group $\psi_{I_1 ... I_n}$ with the Weyl index being suppressed. $\psi$ is a symmetric tensor and satisfies the spinor trace condition
\begin{align}
	\gamma^J \psi_{J I_2 ... I_n} = 0
	\label{eq:spinor_trace}
\end{align} 
with the Dirac matrices $\gamma^I$. (\ref{eq:spinor_trace}) reduces the total number of degrees of freedom in the chiral components of $\psi$ to $ 2\left( n+\frac{1}{2} \right) + 1$ as expected for a spin $n + \frac{1}{2}$ representation\footnote{The number of possible independent values of $\psi$ is given by the number of partitions of n identical blocks into four parts.}.
To run the program of loop quantum gravity for half-integral spin fields, we would need a Lagrangian from which we can derive the generalization of the Dirac equation
\begin{align}
	(-i e^\mu_I \gamma^I \mathcal{D}_\mu + m) \psi_{I_1 ... I_n} &= 0 \label{eq:Rarita_gen} \\
	\eta^{IJ} e^\mu_I \mathcal{D}_\mu \psi_{J I_2 ... I_n} &= 0,
	\label{eq:half_div}
\end{align}
where we carefully intertwine 4-dimensional representation space indices $I, J, ...$ with 4-dimensional spacetime indices $\mu, \nu, ...$ nontrivially via the tetrad $e^\mu_I$ (cf. (\ref{eq:Dirac_act_original})). Unfortunately, (\ref{eq:half_div}) is not an Euler-Lagrange equation, i.e. cannot be derived from an action by a variational principle unless we introduce auxiliary matter fields \cite{Singh:1974rc}. For a spin $\frac{3}{2}$ particle, for instance, the Lagrangian density which yields the equations (\ref{eq:Rarita_gen}) and (\ref{eq:half_div}) for the massless case reads
\begin{align}
	\mathcal{L} &= \frac{\beta}{2} \eta^{IJ} \psi_I \left( i \gamma^K e_K^\mu \mathcal{D}_\mu - m \right) \psi_J - \frac{2}{3} \beta \chi \left( e^\mu_I \eta^{IJ} \mathcal{D}_\mu \psi_J \right) \nonumber \\[1 ex]
	&- \frac{1}{3} \beta \chi \left( i \gamma^K e_K^\mu \mathcal{D}_\mu + 2m \right) \chi,
	\label{eq:3_2_Lagrangian}
\end{align}
where we introduced an auxiliary Dirac spinor field $\chi$. The corresponding Euler-Lagrange equations can be written in the following form \cite{Singh:1974rc}
\begin{align}
	&(-i \gamma^K e_K^\nu \mathcal{D}_\nu + m) \psi^I + \frac{i}{2} \gamma^I e^\mu_K \mathcal{D}_\mu \psi^K = \nonumber \\
	&= \frac{2}{3} \left( e^\mu_J \eta^{IJ} \mathcal{D}_\mu + \frac{1}{4} \gamma^I \gamma^K e^\mu_K \mathcal{D}_\mu \right) \chi
	\label{eq:ELE_1} \\
	&e^\mu_I \mathcal{D}_\mu \psi^I = -(i \gamma^K e_K^\mu \mathcal{D}_\mu + 2m) \chi.
	\label{eq:ELE_2}
\end{align}
The coefficients in front of each of the terms in (\ref{eq:3_2_Lagrangian}) are chosen such that the auxiliary spinor field vanishes on shell. The equation $\chi = 0$ can be obtained by contracting  (\ref{eq:ELE_1}) with $e^\mu_I \mathcal{D}_\mu$ and substituting (\ref{eq:ELE_2}) therein. This reproduces the equations (\ref{eq:Rarita_gen}) and (\ref{eq:half_div}) for the special case of a spin $\frac{3}{2}$ particle. For higher spins, the Lagrangian yielding the equations of motion (\ref{eq:Rarita_gen}, \ref{eq:half_div}) can be constructed in a similar way \cite{Singh:1974rc}. For the spin $\frac{3}{2}$ particle, however there is a way to write the Lagrangian without the necessity of introducing auxiliary matter fields. The Rarita-Schwinger Lagrangian \cite{Rarita_41} takes the compact form
\begin{align}
	\mathcal{L}_\text{RS} &= \bar{\psi}_I \left( \gamma^{IJK} e^\mu_J \mathcal{D}_\mu - i m \sigma^{IK} \right) \psi_K \nonumber \\
	&\text{with } \gamma^{IJK} = \frac{1}{3!} \gamma^{[I} \gamma^J \gamma^{K]} \quad \text{ and } \quad \sigma^{IK} = [\gamma^I, \gamma^K].
\end{align}
In the canonical theory, the Rarita Schwinger field is described by $\Psi_i$ and the corresponding canonical momentum, which is linear in $\overline{\Psi}_0$ \cite{Bodendorfer:2011pb, Eder:2020uff, Eder21}. Again it turns out to be advantageous to go over to density weight $\frac{1}{2}$ fields. On the kinematical level, the four-dimensional Rarita Schwinger field can thus be quantized like a triple of Dirac fields. One can perform the analogue constraint analysis as in the Dirac case and again arrives at contributions to Gauss, diffeomorphism and Hamilton constraint, as well as new constraints coming from the fact that $\Psi_0$ turns out to be non-dynamical.

The classical theory of higher half-integral spin quantum fields can also be systematically constructed similar to (\ref{eq:general_boson_action}). Again, we refrain from writing out the whole Lagrangian, but indicate the dependence of $\mathcal{L}$ on the auxiliary fields \cite{Singh:1974rc}
\begin{align}
	\mathcal{L} &= \frac{\beta}{2} \sqrt{-\det(g)} \psi^{(j)}_{I_1 ... I_n} \left( i e^\mu_I \gamma^I \nabla_\mu - m \right) \psi^{(j) \, I_1 ... I_n} + \nonumber \\
	&+ \mathcal{O}(\psi^{(j-1)})
	\label{eq:general_fermion_action}
\end{align}
with $n = j - \frac{1}{2}$ and the matter degrees of freedom described by the symmetric tensor field $\psi^{(j)}$, which also carries a Dirac spinor representation. Also here, the important term yielding the desired constraints is the first one in (\ref{eq:general_fermion_action}), while the other terms vanish if we implement the on shell conditions. The Gauss and diffeomorphism constraints finally have the same form as (\ref{eq:general_Gauss}) and (\ref{eq:general_diffeo}). The Hamilton constraint on the other hand will have the form
\begin{align}
	H &= e^a_I \phi^{I_1 ... I_n} \gamma^I \mathcal{D}_a \phi_{I_1 ... I_n} + m^2 \phi_{I_1 ... I_n} \phi^{I_1 ... I_n} + \nonumber \\
	&+ \mathcal{O}(\psi^{(j-q)}).
	\label{eq:half_integral_Hamilton}
\end{align}
The Hamilton constraint differs from the integral case by its linear dispersion relation. In the case of half-integral spin, we conjecture that it will be possible to introduce half-densities $\xi$ describing the matter field without changing the symplectic structure. This is harder in the case of integral spin, but an adequate redefinition of the connection also yielded a formulation in terms of half-densities in the spin 0 case.

\section{Electromagnetic Charge and Antiparticles}
\label{sec:elmag}
In this last section, we want to challenge the extension of loop quantum gravity with matter fields by a U(1) gauge field. This is used to describe the electromagnetic charge and interaction among particles as well as between particles and spacetime. Therefore, it yields a better suiting description of most particles of the standard model of particle physics.

On the level of loop quantum gravity without matter fields, there have already been investigations on the loop quantization of a U(1) theory \cite{Corichi97, Corichi:1997us} and also \cite{Thiemann_1998} discussed that case implicitly in the context of spin $\frac{1}{2}$ particles. We want to review these results and combine them with our previous results.

We will start with a formulation of the constraint algebra of the U(1) and SU(2) Gauss constraint as well as the diffeomorphism and Hamilton constraint. Subsequently, we can define creation and annihilation operators for particles and antiparticles analogously to the path observables of \autoref{sec:LQG_matter}.

\subsection{A Kinematical Hilbert Space for Charged Fermions}
We will start with the  Yang-Mills action for a U(1) connection one-form $\elmagA_\mu$ on curved spacetime
\begin{align}
	S_\text{YM} = -\frac{1}{4} \int dt \int_\Sigma d^3 x \sqrt{-\det(g)} \elmagF_{\mu \nu} \elmagF_{\rho \sigma} g^{\mu \rho} g^{\nu \sigma},
	\label{eq:act_U1}
\end{align}
with the curvature $\elmagF_{\mu \nu} =  \partial_{\mu} \elmagA_{\nu} - \partial_{\nu} \elmagA_{\mu}$. In $\elmagF_{\mu \nu}$, we can replace the partial derivatives by covariant derivatives $\mathcal{D}$ corresponding to the Levi-Civita connection $\Gamma$ for the transport of spacetime structures, the Ashtekar connection for SU(2) structures and the U(1) connection $\elmagA_\mu$ for U(1) structures. 

Performing a Legendre transformation of (\ref{eq:act_U1}) with respect to $\elmagA_\mu$ and  the conjugate momentum
\begin{align}
	\elmagE^\mu = \frac{\partial L}{\partial (\partial_0 A_\mu)} = \sqrt{-\det(g)} \elmagF^{\mu 0},
\end{align}
we end up with a Hamiltonian consisting of the following constraints only \cite{Thiemann_1998_QSD}
\begin{align}
	G_i &= \mathcal{D}_a E^a_i + i \left( \xi^\dagger \tau_i \xi + \rho^\dagger \tau_i \rho \right) \\
	\underline{G} &= \mathcal{D}_a \elmagE^a + \left( \xi^\dagger \xi + \rho^\dagger \rho \right) \\
	V_a &= F^i_{ab} E^b_i + \elmagF_{ab} \elmagE^b + \frac{i}{2} \sqrt{\det(q)} \left( \xi^\dagger \mathcal{D}_a \xi + \rho^\dagger \mathcal{D}_a \rho - c.c. \right) \\
	H &= \frac{1}{2 \sqrt{\det(q)}} \Bigg[ \left( 2[K_a, K_b]^i - F^i_{ab} \right) [E^a, E^b]_i + \nonumber \\
	&+ E^a_j \mathcal{D}_a \left( \xi^\dagger \sigma^j \xi + \rho^\dagger \sigma^j \rho \right) + \nonumber \\
	&- i E^a_j \left( \xi^\dagger \sigma^j \mathcal{D}_a \xi - \rho^\dagger \sigma^j \mathcal{D}_a \rho - c.c. \right) + \nonumber \\
	&+ E^a_j K^j_a \left( \xi^\dagger \xi - \rho^\dagger \rho \right) + q_{ab} \left( \elmagE^a \elmagE^b + \elmagB^a \elmagB^b \right) \Bigg].
\end{align}
where we defined $\elmagB^a := \frac{1}{2} \tensor{\epsilon}{^b^c_a} F_{bc}$. The U(1) Gauss constraint corresponds to the Lagrange multiplier $\elmagA_0$. Also the diffeomorphism and Hamilton constraint arise from the action in an analogous manner. This is not surprising, since we can express gravity theory in terms of Ashtekar's variables \cite{Ashtekar_86} as an SU(2) Yang-Mills theory as well. A more elegant, geometric approach to the constraint analysis of the U(1) Yang-Mills theory is derived by \cite{Eder17, Eder18}.

The U(1) Gauss constraint generates U(1) gauge transformations. Note that both, $\xi$ and $\rho$, have the same sign in the U(1) gauge constraint, which corresponds to the same sign of electromagnetic charge. Importantly, the two Gauss constraints decouple, which can be seen by calculating the Poisson brackets between the smeared SU(2) and U(1) Gauss constraints
\begin{align}
	\{G(\Lambda), \underline{G}(\underline{\Lambda})\} = 0.
\end{align}
Also, since $\underline{\Lambda}$ is abelian, one can show
\begin{align}
	\{\underline{G}(\underline{\Lambda}), \underline{G}(\underline{\Lambda}')\} = 0.
\end{align}
As for the vector constraint $V_a$, we can reformulate it in the same way as before yielding the constraint $W(\vec{N})$, which generates spatial diffeomorphisms. This constraint behaves similarly as an element of the Poisson algebra. The commutator with the U(1) Gauss constraint takes the form
\begin{align}
	&\left\{\underline{G}(\underline{\Lambda}), W(\vec{N}) \right\} = \int_\Sigma d^3 x \, \left\{\underline{\Lambda} \mathcal{D}_a \elmagE^a, \left( \leftidx{^\text{sp}}{\mathcal{L}}{_{\vec{N}}} \elmagA \right)_b \elmagE^b \right\} \nonumber \\
	&+ \left\{ \underline{\Lambda} (\xi^\dagger \xi + \rho^\dagger \rho), i N^a (\xi^\dagger \partial_a \xi + \rho^\dagger \partial_a \rho - c.c.) \right\} \nonumber \\
	&= \underline{G}\left( \vec{N}(\underline{\Lambda}) \right) = \underline{G}\left( \leftidx{^\text{sp}}{\mathcal{L}}{_{\vec{N}}} \underline{\Lambda} \right),
\end{align}
which is analogue to $\{G(\Lambda), W(\vec{N})\} = G(\leftidx{^\text{sp}}{\mathcal{L}}{_{\vec{N}}} \Lambda)$ \cite{thiemann_2007}. The Poisson relations excluding the U(1) Gauss constraint $\underline{G}(\underline{\Lambda})$ remain untouched. This ensures that the kinematical Hilbert space with the two Gauss constraints and the diffeomorphism constraint being implemented is stable under the action of the constraint operators, since the constraint algebra is closed under the Poisson bracket.

Let us also review the unconstrained Hilbert space of the theory including U(1) representations \cite{Corichi97}. We build the Hilbert space additionally with U(1) holonomies
\begin{align}
	\elmagh{e} (\elmagA) := \mathcal{P} \exp \int_e \elmagA \quad \in \text{U(1)},
\end{align}
which satisfies the same properties as the SU(2) holonomies in addition to the fact that they are abelian. These will build up our Hilbert space, as we can formulate a $^*$ algebra with them together with the electric fields $E$ and $\elmagE$ \cite{Corichi97}. 

Hence, we define the Hilbert space of charged matter fields on loop quantum gravity as the Cauchy completion of the span of smooth cylindrical functions with respect to both connections $A$ and $\elmagA$ 
\begin{align}
	\mathcal{H} &:= \bigoplus_\gamma \mathcal{H}_\gamma ' \nonumber \\
	\mathcal{H}_\gamma &:= \overline{ \Spann\left\{ \Psi_\gamma [A, \elmagA] \in \Cyl^\infty_\gamma \right\}},
\end{align}
where we now call a function of $A$ and $\elmagA$ cylindrical if it can be written as a function of holonomies
\begin{align}
	\Psi[A, \elmagA] = f(h_{e_1}, ..., h_{e_N}, \elmagh{e_1}, ..., \elmagh{e_N}) \quad \in \text{U(1)}.
\end{align}
The prime $'$ in $\mathcal{H}'_\gamma$ denotes that we only consider the orthogonal components of each of the $\mathcal{H}_\gamma$, i.e. $\mathcal{H}_\gamma '$ does not include cylindrical functions which are also cylindrical with respect to a graph $\tilde{\gamma} \subsetneq \gamma$ which is strictly included in $\gamma$. 

The inner product can be constructed just like the Ashtekar-Lewandowski inner product but using the Haar measure of SU(2)$\times$U(1) instead of SU(2). The orthogonal components are further decomposed into the irreducible representations of SU(2) labelled by spin quantum numbers $j \in \frac{\mathds{N}_0}{2}$ and U(1) labelled by charge quantum numbers $n \in \mathds{Z}$. This yields the decomposition
\begin{align}
	\mathcal{H}_\gamma ' \cong \bigoplus_{\vec{j}, \vec{l}, \vec{n}} \mathcal{H}_{\gamma, \vec{j}, \vec{l}, \vec{n}}',
\end{align}
where the spin vectors $\vec{j}$ and $\vec{l}$ encode the spin representations of the SU(2) holonomies and $\vec{n}$ encodes the representation of the U(1) holonomies and the $\mathcal{H}_{\gamma, \vec{j}, \vec{l}, \vec{n}}'$ only contain cylindrical functions admitting the respective spin or charge representations of the SU(2) and U(1) holonomies. 

In order to solve the U(1) Gauss constraint, note that the $n$ representation of $\elmagh{e}$ transforms like a holonomy
\begin{align}
	\jrep{n}{\elmagh{e}} &\mapsto \jrep{n}{g(e(0))} \jrep{n}{\elmagh{e}} \jrep{n}{g^{-1}(e(1))} \nonumber \\
	&= e^{-in \phi(e(0))} \jrep{n}{\elmagh{e}} e^{in \phi(e(1))},
	\label{eq:U1_trafo_h}
\end{align}
with $\phi: \Sigma \to \mathfrak{u}(1)$. Taking into account the matter fields, we find the transformation
\begin{align}
	\theta_p \mapsto e^{-i n_\theta \phi(p)} \theta_p.
	\label{eq:U1_trafo_t}
\end{align}
If we consider gauge invariant cylindrical functions only, all the exponents in the gauge transformation of (\ref{eq:U1_trafo_h} - \ref{eq:U1_trafo_t}) have to cancel. The U(1) invariance can hence be translated into the simple condition
\begin{align}
	\sum_{\substack{e \in E(v) \\ e(1) = v }} n_e - \sum_{\substack{ e \in E(v) \\ e(0) = v}} n_e - \sum_{\theta \in \Theta(v)} n_\theta = 0 \qquad \forall v \in V(\gamma),
	\label{eq:U1_invariance}
\end{align}
where $E(v)$ denotes the set of edges which start or end at the vertex $v \in V(\gamma)$ and $\Theta(v)$ the set of matter fields which are attached to $v$. Note that $n_\theta$ is fixed for a fixed particle type $\theta$. In particular for spin $\frac{1}{2}$ fermions, it holds that $n_\theta = n_\omega$ as pointed out before. All the quantum states which satisfy (\ref{eq:U1_invariance}) in addition to being SU(2) invariant define the Hilbert space $\mathcal{H}_G$ of gauge invariant quantum states. However, for U(1) invariance we need the notion of antiparticles, which is not yet present in our discussion. Let us consider the following U(1) \textit{variant} example
\begin{example}
	\label{ex:charged_pair}
	Consider two charged fermions of spin $\frac{1}{2}$ and charge $n = 1$ connected by an edge $e$
	\begin{align}
		&\hat{\theta}_p^A \epsilon_{AB} \tensor{\jrep{\frac{1}{2}}{h_e}}{^B_C} \jrep{1}{\elmagh{e}} \hat{\theta}_q^C \vac \nonumber \\[2 ex]
		&= \quad \adjustbox{valign = c}{
			\begin{overpic}[scale = 0.15]{Singlet_asymmetric.pdf}
				\put(37, 22){\scriptsize $j = \frac{1}{2}$}
				\put(65, 22){\scriptsize $n = 1$}
				\put(-10, 12){\scriptsize $n = 2$}
				\put(7, -3){\scriptsize $j = 0$}
				\put(30, 10){\scriptsize $p$}
				\put(100, 10){\scriptsize $q$}
			\end{overpic}
		}_.
	\end{align}
	Since the two particles carry the same charge, a total electromagnetic charge of 2 quanta is flowing out of the two fermion system.
\end{example}
The particles $\theta$ carry an intrinsic charge and therefore can be seen as sources (or sinks for the opposite sign, respectively) of electromagnetic charge. The U(1) holonomies on the other hand, do not provide a source of charge because the same amount of electromagnetic charge flows through every point along the holonomy. 

For U(1) invariant states, the U(1) Gauss constraint states that the overall electromagnetic flux of the quantum state vanishes. Therefore, there has to be either vacuum or both, positively and negatively charged particles in order to balance each other. This raises the question of how to describe antiparticles (i.e. particles of the same type with an electromagnetic charge of opposite sign) within our theory.

\subsection{Antiparticles}
\label{subsec:anti}
From the classical action, the concept of antiparticles is not manifest. Indeed, also in flat spacetimes, antiparticles appear first in the Fourier transform of the quantum fields. In loop quantum gravity, however, we cannot access this tool, but may define an analogue version of the particle-antiparticle pair inspired by flat quantum field theory
\begin{align}
	\hat{\theta} = \frac{1}{\sqrt{2}} \left( \hat{\theta}_+ + \hat{\pi}_{\theta_-} \right) \label{eq:anti_cre} \\
	\hat{\pi}_{\theta} = \frac{1}{\sqrt{2}} \left( \hat{\theta}_- + \hat{\pi}_{\theta_+} \right),
	\label{eq:anti_cre_2}
\end{align}
and similar for $\omega$. Here, we defined the creation $\hat{\theta}_\pm$ and annihilation operators $\theta^\dagger = -i \hat{\pi}_{\theta_\pm}$ for particles ($+$) and antiparticles ($-$).

The chiral components of the Dirac spinor are both, creating a particle, but also annihilating an antiparticle if we are serious about the analogy in flat quantum field theory. The momenta $\pi_\theta$ and $\pi_\omega$ on the other hand should also create an antiparticle in addition to annihilating a particle. The definitions (\ref{eq:anti_cre}) and (\ref{eq:anti_cre_2}) are moreover compatible with the reality conditions $\pi_{\theta} = i \theta^\dagger$ and impose the new reality conditions
\begin{align}
	\pi_{\theta_\pm} = i \theta_\pm^\dagger.
\end{align}
Another way of recovering particle and antiparticle as Weyl spinors in the theory is achieved by switching the terms $\theta$ and $\theta^\dagger$ for one of the two Weyl components \cite{Thiemann_private}. This way, also the electromagnetic charge of one of the chiral components is flipped. This does not change the anticommutation relations and is therefore a freedom of choice in the quantum theory. Let us finally look at the U(1) invariant version of example \ref{ex:charged_pair}.
\begin{example}
	\label{ex:particle_antiparticle}
	Consider the particle-antiparticle pair connected by an SU(2) and a U(1) holonomy
	\begin{align}
		&\hat{\theta}^{\dagger \, p}_{A} \tensor{\jrep{\frac{1}{2}}{h_e}}{^A_B} \jrep{1}{\elmagh{e}} \hat{\theta}_{q}^B \vac \nonumber \\[1 ex]
		&= \hat{\theta}^p_{- \, A} \tensor{\jrep{\frac{1}{2}}{h_e}}{^A_B} \jrep{1}{\elmagh{e}} \hat{\theta}_{+ \, q}^B \vac \nonumber \\[2 ex]
		&=\quad\adjustbox{valign = c}{
			\begin{overpic}[scale = 0.15]{Baez_pi_theta.pdf}
				\put(25, 14){\scriptsize $j = \frac{1}{2}$}
				\put(55, 14){\scriptsize $n = 1$}
				\put(-7, 0){\scriptsize $p$}
				\put(103, 0){\scriptsize $q$}
			\end{overpic}
		}_.
	\end{align}
	The antiparticle is depicted by an empty circle. Note that the former annihilation operator $\hat{\theta^\dagger}$ does not annihilate the vacuum anymore, but the operators $\theta^\dagger_\pm$ do. Because of that, only the creation operators $\hat{\theta}_\pm$ survive when acting on the vacuum state. Moreover, since we have both, a source and a sink of electromagnetic charge, the U(1) Gauss constraint is satisfied.
\end{example}
Example \ref{ex:particle_antiparticle} shows that we can define the path observables creating elements of gauge invariant quantum states also for SU(2)$\times$U(1) gauge theory. The path observable in example \ref{ex:charged_pair} carries a total charge of 2 quanta and thus cannot be made U(1) invariant without introducing extra structure. Instead, we would have to add a second pair of antiparticles, which absorbs the excess charge. This might look as follows
\begin{example}
	Consider a pair of particles and a pair of antiparticles each connected by an edge with spin $\frac{1}{2}$ and charge $1$. Furthermore, the two pairs are connected by another edge, which does not carry any spin but neutralizes the charge of the two pairs. The pictorial representation reads
	\begin{align}
		\adjustbox{valign = c}{
			\begin{overpic}[scale = 0.15]{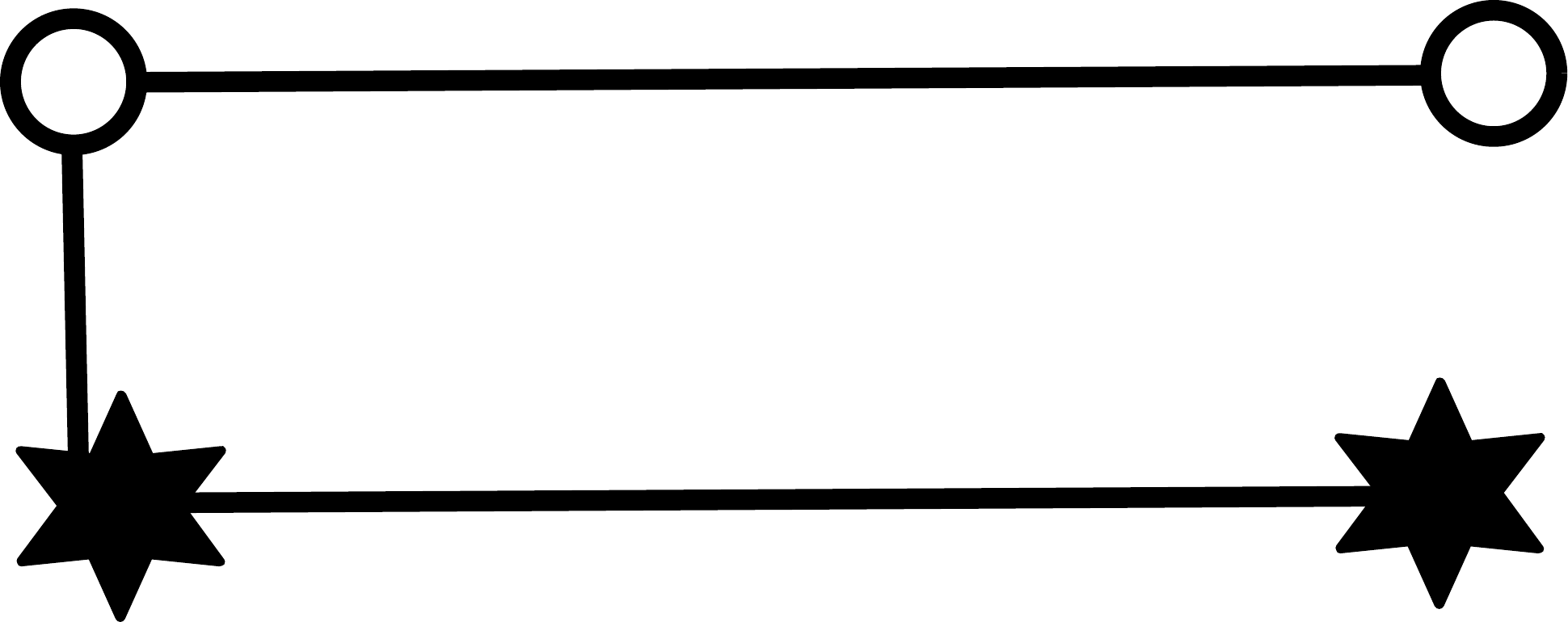}
				\put(25, -2){\scriptsize $j = \frac{1}{2}$}
				\put(55, -2){\scriptsize $n = 1$}
				\put(25, 40){\scriptsize $j = \frac{1}{2}$}
				\put(55, 40){\scriptsize $n = 1$}
				\put(-23, 20){\scriptsize $j = 0$}
				\put(-23, 12){\scriptsize $n = 2$}
				\put(-7, 0){\scriptsize $p$}
				\put(100, 0){\scriptsize $q$}
				\put(-7, 30){\scriptsize $r$}
				\put(100, 30){\scriptsize $s$}
			\end{overpic}
		}_.
	\end{align}
	\label{ex:U1_two_pairs}
\end{example}
Example \ref{ex:U1_two_pairs} can be further generalized using an arbitrary gauge invariant generalized spin network state the same way as in the uncharged case. The fact that we need as many particles as antiparticles (or equivalently positive and negative charges) is also in agreement with flat quantum field theory.

\section{Summary and Outlook}
In this work, we have studied various kinematical aspects of the coupling of quantum matter to loop quantum gravity. 

Following \cite{Thiemann_1998, Thiemann_1998_QSD}, but using the language of Fock spaces, we defined the Hilbert space $\mathcal{H}$ and basic operators for a quantum theory of spin $\frac{1}{2}$ fields, and we generalized it to higher spin quantum fields in a straightforward way. 
An orthonormal basis for the gauge invariant states in $\mathcal{H}$ can be found by the generalization of the notion of spin network states known from matter-free loop quantum gravity. This generalization contains matter fields which can be located at the vertices of the underlying spin network graph $\gamma$. Inspired by the gauge invariant observables defined by \cite{MoralesTecotl:1994ns,Baez:1997bw}, we were able to introduce a closed (anti)commutator algebra of creation and annihilation operators for gauge invariant generalized spin network states.\\

While the Gauss constraint could be directly understood to ensure the gauge invariance of the states, one has to add a phase space dependent generator of a gauge transformation to the diffeomorphism constraint before it generates spatial deformations without gauge transformations (Appendix \ref{app:spatial_diffeo}). We have checked that this is possible at least locally, which is enough for the further considerations in the present work. It also fixes the order of implementation of the constraints, since the action of the diffeomorphism constraint would generate terms proportional to the Gauss constraint otherwise. 

The implementation of the diffeomorphism constraint on the space of gauge invariant generalized spin network states was performed with the group averaging technique in analogy to the matter-free case \cite{Ashtekar_1995, Ashtekar_2004}. We paid special attention to the interplay between the (anti)symmetrization imposed on quantum states that are based on graphs with symmetries by the diffeomorphism constraint on the one hand, and the (anti)symmetrization of the state due to the statistics of the matter field on the other. The fact that (anti)symmetry is imposed in some cases due to the diffeomorphism constraint is a novel feature in loop quantum gravity. In special symmetric cases and by consideration of particle exchange generated by a graph symmetry, we have found simple rules which characterize whether a state is annihilated by the diffeomorphism averaging or not. For these considerations, we assumed a spin-statistics connection motivated by the fact that a spin-statistics theorem exists for quantum field theories on curved spacetime \cite{Verch:2001bv}. 

While we found that it is straightforward to generalize the quantum kinematics of  \cite{Thiemann_1998,Thiemann_1998} and \cite{Bojowald_2008} to matter theories with higher spin, it is an important question whether this quantum theory can be derived from a classical higher spin action. These actions contain auxiliary matter fields which have no physical significance whatsoever as they vanish on shell. Using this technique, the kinematical constraints on the classical phase space could be elaborated and indeed take a similar form as in the lower spin case. 

Although we completely discussed the case of spin 0 and $\frac{1}{2}$ only, we can conjecture a general behavior by solving the on shell conditions for the auxiliary fields in the higher spin case. However, it is not clear how to consistently solve the second class constraint with the Lagrange multiplier $\leftidx{^-}{A}{_a}$ in order to get from the SO(3, 1) connection $\omega$ to an SU(2) connection $A$. In addition, in the four-dimensional representation of the spin 1 particle we encountered further second class constraints, since the 0-component of the momentum $\pi^0$ seems not to be an independent variable. For the simplified situation of flat spacetime, solving the additional constraints yields nonlocal dependencies, and hence a complicated theory. In any case, this should be further investigated before tackling even higher spins. 

Finally, we briefly considered charged Dirac fermions coupled to gravity and U(1) Yang-Mills theory, and rederived earlier results \cite{Corichi:1997us,Thiemann_1998_QSD}. We found a way to represent both, particle and anti-particle on the Fock space. Due to the requirement of gauge invariance, creation of particles goes hand in hand with that of anti-particles. We have sketched the consequences for the gauge invariant observables introduced earlier. We note that it would not be easy to tackle the classical theory of electromagnetically charged higher spin fields. The problem here is that it is not clear whether the vanishing conditions for the auxiliary fields still hold. There is already existing work on a Lagrangian formulation of higher spin theories including an electromagnetic field \cite{Fiertz_39, Vasiliev:2004qz}, which might be used to extend the available theory to gravitationally and electromagnetically interacting theories.

Altogether we found that there is a very simple and elegant kinematics for coupling quantum matter of arbitrary spin to loop quantum gravity. Implementation of the Gauss constraint leads to a tight coupling between matter and gravity already at the kinematical level, and the diffeomorphism constraint imposes interesting symmetry properties on the joint quantum states in certain situations. But obviously there are still many open problems on this path to a unified theory of matter and gravity.

\begin{acknowledgments}
    We thank Kristina Giesel and Thomas Thiemann for interesting discussions during the completion of this work. Moreover, we thank Thomas Thiemann for very detailed and valuable feedback.   
	
	We thank the anonymous referee for a suggestion that improved the presentation of the material substantially.
	
    RM thanks the Elite Graduate Programme of the Friedrich-Alexander-Universität Erlangen-Nürnberg for its support and for encouraging discussions among its members and faculty.
    
    H.S. would like to acknowledge the contribution of the COST Action CA18108.
\end{acknowledgments}

\appendix
\section{The Generator of Purely Spatial Transformations}
\label{app:spatial_diffeo}
From the combined action of Holst gravity \cite{Holst_96} with Dirac fermions we can read off the smeared diffeomorphism constraint \cite{Thiemann_1998_QSD},
\begin{align}
	V(\vec{N}) &= \int_{\Sigma} \text{d} x^3 N^a V_a \nonumber \\ 
	&= \int_{\Sigma}\text{d} x^3 N^a \Big( \frac{1}{\kappa} E_i^b F^i_{ab} + \nonumber \\
	&+ \frac{i}{2} \left( \xi^\dagger \mathcal{D}_a \xi + \rho^\dagger \mathcal{D}_a \rho - c.c. \right) \Big),
	\label{constraint_V}
\end{align}
where $\xi$ and $\rho$ are left- and right-handed half-density spinors in the Weyl representation as defined in \cite{Thiemann_1998} and $\mathcal{D}$ the SU(2) covariant derivative. However, following the standard discussions of the vacuum theory \cite{thiemann_2007, Ashtekar_2004}, we do not stick to (\ref{constraint_V}) as a diffeomorphism constraint. We will rather add a term proportional to the Gauss constraint, with which we might define a new constraint within the same constraint algebra. Using integration by parts and vanishing boundary conditions, we find the following form of the new constraint:
\begin{align}
	W(\vec{N}) &= \int_{\Sigma} \text{d} x^3 \left( N^a V_a - N^a A_a^i G_i \right) \nonumber \\
	&= \int_{\Sigma} \text{d} x^3 \Big( E_i^a \left( \partial_a N^b A_b^i + N^b \partial_b A_{a}^i \right) + \nonumber \\
	&+ \frac{i}{2} N^a \left( \xi^\dagger \partial_a \xi + \rho^\dagger \partial_a \rho - c.c. \right) \Big).
	\label{constraint_W}
\end{align}
The constraint $W(\vec{N})$ is what is often called the diffeomorphism constraint in the literature \cite{thiemann_2007, alej2004introduction, giesel2017kinematical}. The reason for this name becomes apparent when calculating the action of (\ref{constraint_W}) on the phase space variables. Furthermore, it is important to note that the constraint $W(\vec{N})$ is dependent on a section in the principal fiber bundle, which is in general only defined locally. Hence, we cannot subtract the terms encoding the gauge transformations for \textit{any} diffeomorphism but only on those with a support in an open neighborhood around a given point\footnote{As a matter of fact, this might yield nontrivial effects depending on the topology of $\Sigma$.}. The Poisson brackets of the constraint with respect to the phase space coordinates $\left( E^a_i, \xi, \rho, A_a^i, i \xi^\dagger, i\rho^\dagger \right)$ read
\begin{align}
	\left\{ W(\vec{N}), E_i^a \right\} &= N^b \partial_b E_{i}^a - \partial_b N^a E^b_i + \partial_b N^b E_i^a \nonumber \\
	&= \left( \leftidx{^\text{sp}}{\mathcal{L}_{\vec{N}}}{} E \right)^a_i \label{Poisson_W_E} \\
	\left\{ W(\vec{N}), A_a^i \right\} &= \partial_a N^b A_b^i + N^b \partial_b A^i_a \nonumber \\
	&= \left( \leftidx{^\text{sp}}{\mathcal{L}_{\vec{N}}}{} A \right)_a^i \\
	\left\{ W(\vec{N}), \xi \right\} &= N^a \partial_a \xi + \frac{1}{2} \partial_a N^a \xi \nonumber \\
	&= \leftidx{^\text{sp}}{\mathcal{L}_{\vec{N}}}{} \xi \\
	\left\{ W(\vec{N}), i \xi^\dagger \right\} &= N^a \partial_a \left( i\xi^\dagger \right) + \frac{1}{2} \partial_a N^a \left(i \xi^\dagger \right) \nonumber \\
	&= \leftidx{^\text{sp}}{\mathcal{L}_{\vec{N}}}{} i \xi \label{Poisson_W_td}
\end{align}
and analogously for $\rho$ and $\rho^\dagger$. The right hand side of (\ref{Poisson_W_E} -- \ref{Poisson_W_td}) describes the infinitesimal action of a finite spatial diffeomorphism, which is in particular the flow of the vector field $\vec{N}$. This is why we call the right hand side of (\ref{Poisson_W_E} -- \ref{Poisson_W_td}) the Lie derivative \cite{Bojowald_2008, thiemann_2007}. The Lie derivative has been chosen to act trivially in the local trivialization chosen for the connection and the fields like $E$ in associated bundles. In the chosen trivializations, the constraint $W(\vec{N})$ thus "ignores" the internal structure of the phase space variables completely.

For the original vector constraint $V(\vec{N})$ we find the following Poisson relations
\begin{align}
	\left\{ V(\vec{N}), E_i^a \right\} &= \tensor[^{\text{sp}}]{\mathcal{L}}{_{\vec{N}}} E_i^a + \tensor{\epsilon}{_i_j^k} N^b A_b^j E_k^a - N^a G_i \nonumber \\
	&= \mathcal{L}_{\vec{N}} E^a_i - N^a G_i \label{Poisson_V_E} \\
	\left\{ V(\vec{N}), A_a^i \right\} &= \tensor[^{\text{sp}}]{\mathcal{L}}{_{\vec{N}}} A_a^i + \mathcal{D}_a \left( N^b A_b^i \right) \nonumber \\
	&= \mathcal{L}_{\vec{N}} A_a^i \label{Poisson_V_A} \\
	\left\{ V(\vec{N}), \xi \right\} &= \tensor[^{\text{sp}}]{\mathcal{L}}{_{\vec{N}}} \xi + N^b A_b^j \tau_j \xi \nonumber \\
	&= \mathcal{L}_{\vec{N}} \xi \label{Poisson_V_t} \\
	\left\{ V(\vec{N}), i \xi^\dagger \right\} &= \tensor[^{\text{sp}}]{\mathcal{L}}{_{\vec{N}}} \left( i \xi^\dagger \right) + N^b A_b^j \tau_j \left( i \xi^\dagger \right) \nonumber \\
	&= \mathcal{L}_{\vec{N}} \left( i \xi^\dagger \right). \label{Poisson_V_td}
\end{align}
We see that $V(\vec{N})$ gives the desired Poisson relations which are gauge covariant Lie derivatives of the canonical variables except in (\ref{Poisson_V_E}). Here, we have a term left which is proportional to the Gauss constraint. From a classical point of view, at least all the phase space variables which lie on the constraint hypersurface are transformed by $V(\vec{N})$ as expected for a gauge covariant diffeomorphism. If we consider the complete phase space a priori, then we will see that $E_i^a$ is transformed like a spacetime vector density and a gauge covector plus a term which is proportional to its covariant divergence.

We conclude by pointing out the difference of the actions of the spatial diffeomorphism constraint $W(\vec{N})$ and the gauge covariant diffeomorphism constraint $V(\vec{N})$. While $W(\vec{N})$ generates the action (via pullback) of diffeomorphisms of $\Sigma$ on phase space functions (see (\ref{Poisson_W_E} - \ref{Poisson_W_td})), the action of $V(\vec{N})$ is more subtle and involves changes in the internal space. In particular, since it is gauge invariant in itself, its action on holonomies can not change the position of their endpoints. Rather it will generate diffeomorphisms that leave these endpoints fixed. Since the constraints $V(\vec{N})$ and $W(\vec{N})$ only differ by a term proportional to $G_i$, they generate the same closed constraint algebra. This becomes apparent when looking at the Poisson relations involving $G(\Lambda)$ and $W(\vec{N})$, for instance
\begin{align}
	\{G(\Lambda), G(\Lambda')\} &= G([\Lambda, \Lambda']) \\
	\{G(\Lambda), W(\vec{N})\} &= -G(\tensor[^{\text{sp}}]{\mathcal{L}}{_{\vec{N}}} \Lambda) \\
	\{W(\vec{N}), W(\vec{M}))\} &= W([\vec{N}, \vec{M}]).
\end{align}
The Poisson brackets involving $G(\Lambda)$ and $V(\vec{N})$ are also closed\footnote{We also stay with a closed Poisson algebra when taking into consideration the Hamilton constraint $H(N)$.}. These take the same form as in matter-free loop quantum gravity \cite{Ashtekar_86, Ashtekar:1987gu}. Therefore, in the quantum theory we will implement the symmetry group $Diff(M)$ consisting of spatial diffeomorphisms only.

\section{Diffeomorphism Symmetric General Spin Network States with Graph Symmetry}
\label{app:General_SNS}
In this appendix, we want to study the behavior of generic spin network states which admit two particles and a graph symmetry which exchanges the particles. The example of section \ref{subsec:specific_SNS} can be generalized by coupling the two spin $\frac{1}{2}$ holonomies to spin $1$ and by closing the graph gauge invariantly. We get
\begin{align}
	\Psi_\text{symm,1} &= \theta^A_p \epsilon_{AB} \tensor{\jrep{\frac{1}{2}}{h_{e_1}}}{^B_C} \tensor{(\sigma_r)}{_i^C^D} \times \nonumber \\
	&\times \tensor{\jrep{\frac{1}{2}}{h_{e_2}}}{^E_D} \epsilon_{EF} \theta^F_q F^i \nonumber \\[2 ex]
	&= \quad \underset{1}{\stackrel{\frac{1}{2} \qquad \qquad \frac{1}{2}}{\leftidx{^p}{\includegraphics[valign=c, scale=0.12]{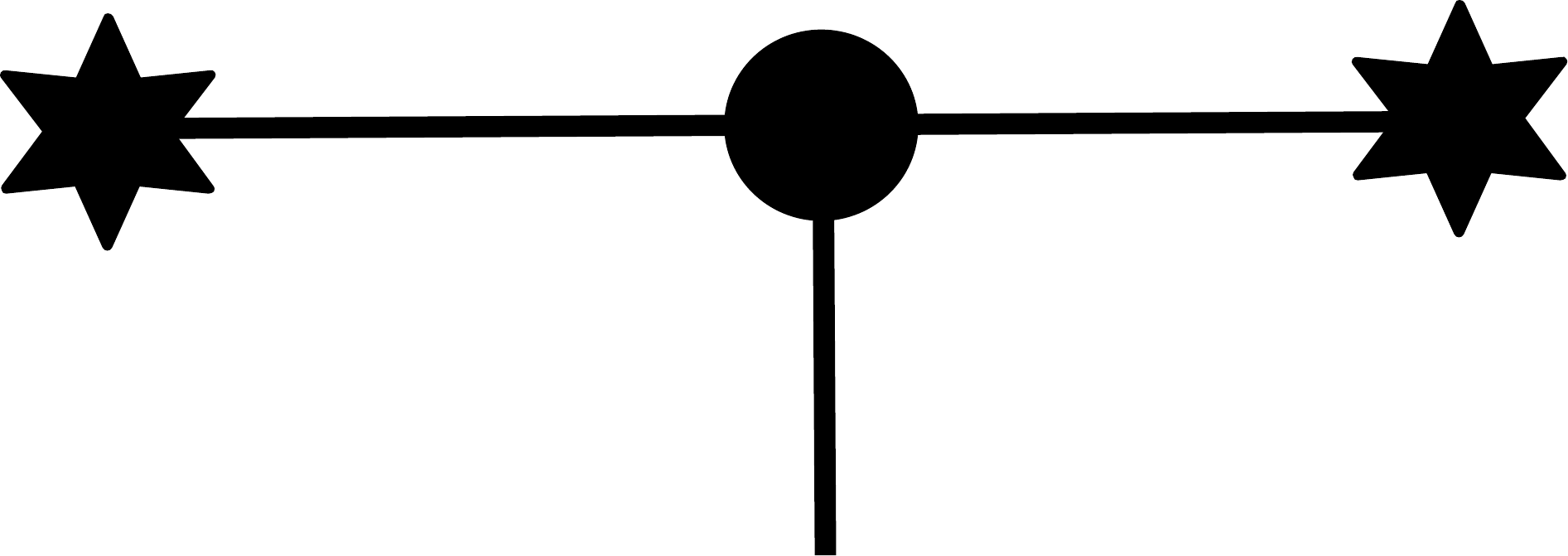}}{^{q}_,}}} \label{eq:two_spin_with_vertex}
\end{align}
where $(\sigma_r)_i$ are the Pauli matrices, which couple $\frac{1}{2} \otimes \frac{1}{2} \otimes 1$ at $r \in \Sigma$, and $F^i$ denotes an arbitrary spin network graph, which couples the rest gauge invariantly, and is invariant with respect to a diffeomorphism $\phi_1$, which rotates $p$ to $q$ and $h_{e_1}$ to $h_{e_2}$ and vice versa. This state is the loop quantum gravity analogue of the triplet state in flat quantum field theory \cite{Mansuroglu:2020acg}. If we act with the graph symmetry $\phi$, we get
\begin{equation}
	\begin{split}
		\hat{U}_{\phi} \Psi_{\text{symm,1}} &= \theta^A_q \epsilon_{AB} \tensor{\jrep{\frac{1}{2}}{h_{e_2}}}{^B_C} \tensor{(\sigma_r)}{_i^C^D} \times \\ &\qquad\qquad\quad \times \tensor{\jrep{\frac{1}{2}}{h_{e_1}}}{^E_D} \epsilon_{EF} \theta^F_p F^i \nonumber \\
		&= - \Psi_\text{symm,1},
	\end{split}    
\end{equation}
where we used the antisymmetry of the two Levi-Civita symbols $\epsilon_{AB}$ and the symmetry of $\sigma^{AB}$ and anti-commuted the two fermions. We deduce that this state will not survive the group averaging procedure, if we stick to the spin-statistics connection. If there is a triplet state in loop quantum gravity, it better not have a graph symmetry as described above.

From the previous calculations, it becomes clear that given a spin network graph which has a suitable graph symmetry the resulting sign is determined by the edge or vertex, respectively, which intersects the symmetry axis. Let us sketch the most general such spin network graph including two particles
\begin{align}
	\Psi_\text{general} = \underset{e_1 \quad \qquad ... \qquad \qquad e_n}{\stackrel{\iota_1 \qquad ... \quad \qquad \iota_{n-1}}{ \leftidx{}{\includegraphics[valign=c, scale=0.05]{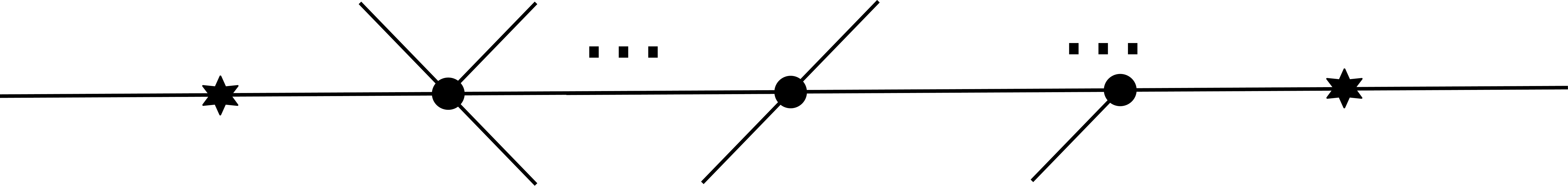}}{}}}_.
	\label{eq:general_symmetric}
\end{align}
The graph symmetry $\phi_\text{general}$ will exchange the two particles and map the edge $e_i$ to $e_{n-i+1}$ and vice versa as well as the points where the intertwiner $\iota_i$ lies to $\iota_{n-i}$ and vice versa. In principle, there is also the possibility that $e_i$ is mapped to $e_{n-i}^{-1}$ (and vice versa) for some $i$. This case is present when the direction of these edges is flipped by $\phi_\text{general}$, which can, however, be transformed to the previous case by inserting two intertwiners $\iota$
\begin{align}
	\tensor{\jrep{j}{h_e}}{^{\mathcal{A}}_{\mathcal{B}}} = (-1)^{2j} \iota_{{\mathcal{B}} {\mathcal{C}}} \tensor{\jrep{j}{h_e^{-1}}}{^{\mathcal{C}}_{\mathcal{D}}} \iota^{{\mathcal{D}} {\mathcal{A}}}.
	\label{eq:edge_direction}
\end{align}
The sign of (\ref{eq:edge_direction}) appears twice as long as $e_{n-i} \neq e_i$ and similar for the contribution of the intertwiner. Therefore, the only contribution to the sign may arise from the object attached to the center of the graph. If $n$ is odd, $e_{\frac{n+1}{2}}$ determines whether the state will survive the group averaging, whereas for $n$ even, $\iota_{\frac{n}{2}}$ determines whether the state will survive the group averaging. 

Let us discuss the case for odd $n$ first. With (\ref{eq:edge_direction}), we can prepare the state such that we only have to consider how to resolve $e_{\frac{n+1}{2}} \to e_{\frac{n+1}{2}}^{-1}$. By preparing the direction of the edges symmetrically around $e_{\frac{n+1}{2}}$, there has to be one nontrivial 2-valent intertwiner $\iota_{{\mathcal{A}} {\mathcal{B}}}$, which takes care of the direction of the edges, at the beginning- or at the endpoint, but not at both points of the holonomy $\jrep{j}{h_{e_{\frac{n+1}{2}}}}$. For instance, the edges $e_{\frac{n-1}{2}}, e_{\frac{n+1}{2}}$ and $e_{\frac{n+3}{2}}$ may take the form
\begin{align}
	\leftidx{}{\includegraphics[valign=c, scale=0.07]{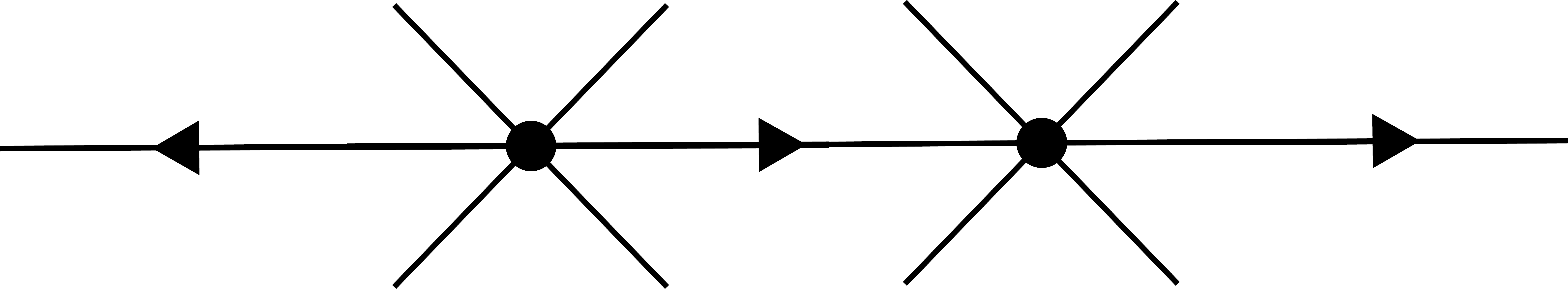}}{}_,
\end{align}
which needs an intertwiner $\iota_{{\mathcal{A}} {\mathcal{B}}}$ at the left vertex but none at the right one. Alternatively, one can invert the directions of $e_{\frac{n-1}{2}}$ and $e_{\frac{n+3}{2}}$ and get an intertwiner $\iota^{{\mathcal{A}} {\mathcal{B}}}$ at the right vertex. The same argument holds for every other edge, too. We can use (\ref{eq:edge_direction}) again, to reformulate
\begin{align}
	\iota_{{\mathcal{A}} {\mathcal{C}}} \tensor{\jrep{j}{h_{e_{\frac{n+1}{2}}}}}{^{\mathcal{C}}_{\mathcal{B}}} &\mapsto \iota_{{\mathcal{A}} {\mathcal{C}}} \tensor{\jrep{j}{h_{e_{\frac{n+1}{2}}}^{-1}}}{^{\mathcal{C}}_{\mathcal{B}}} \nonumber \\
	&= (-1)^{2j} \iota_{{\mathcal{B}} {\mathcal{C}}} \tensor{\jrep{j}{h_{e_{\frac{n+1}{2}}}}}{^{\mathcal{C}}_{\mathcal{A}}}.
\end{align}
Unfortunately, we cannot control what spin couples to the many intertwiners inbetween the two particles. Hence, the spin $j$ of the holonomy lying in the center and the corresponding arising sign is not determined by the spin of the particles $\theta$. That way, the spin of the holonomy intersecting the symmetry axis of $\phi_\text{general}$ has to be of the same type (half-integral or integral) as the spin of the particles $\theta$.

If we have an intertwiner intersecting the symmetry axis, the result is even more subtle. Let us sketch the most general such intertwiner \\
\begin{align}
	\underset{k \qquad \qquad \qquad k}{\stackrel{j \qquad \qquad \qquad j}{\leftidx{}{\includegraphics[valign=c, scale=0.1]{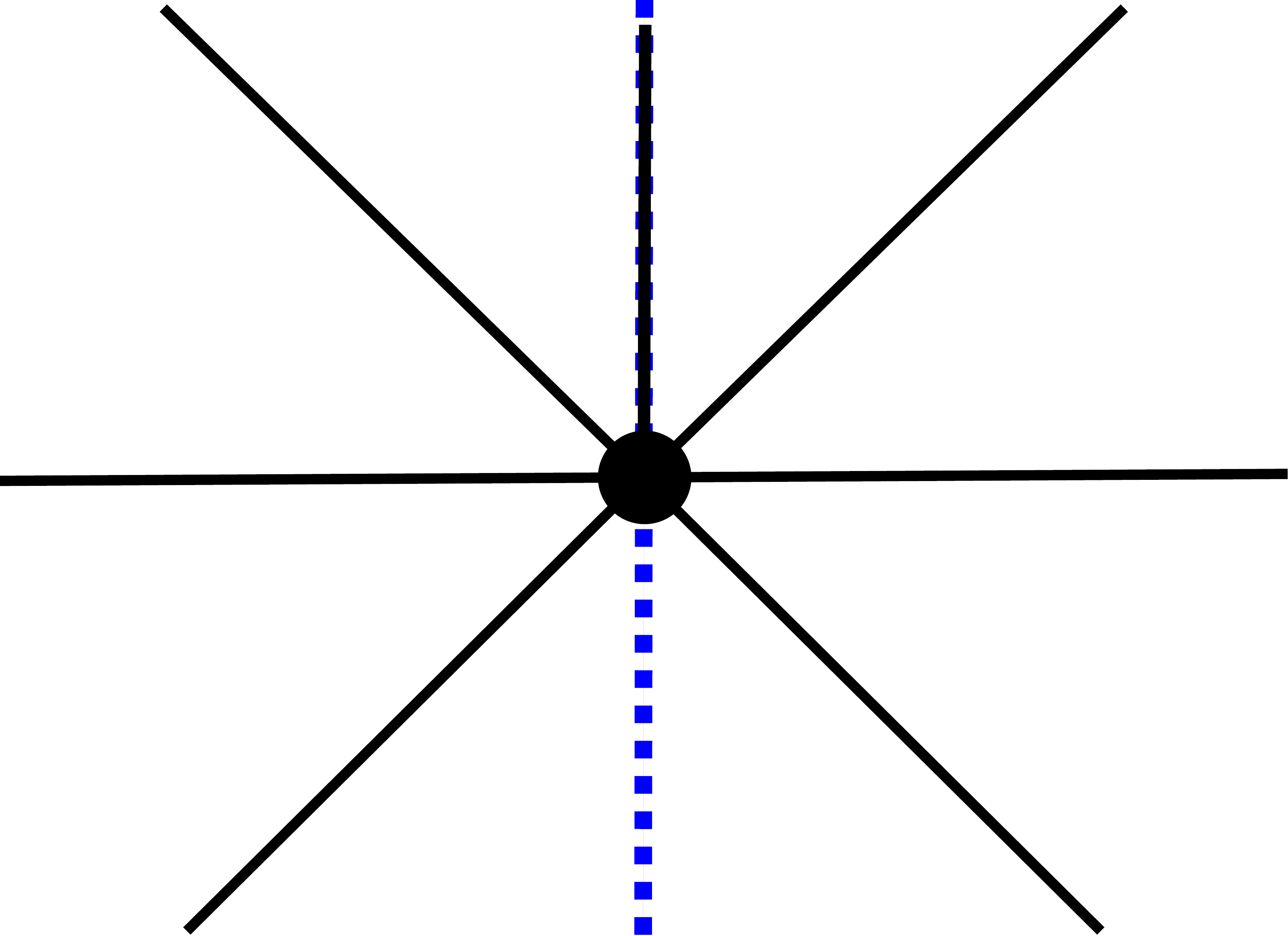}}{}}}_,
	\label{eq:intertwiner}
\end{align}
where we depicted the rotation symmetry axis by a dashed blue line. As we can see, holonomies which lie on the opposite of the symmetry axis must carry the same spin. Apart from that, holonomies might also lie on the symmetry axis. Their spin is arbitrary and does not underlie symmetry constraints, since they are invariant under $\phi_\text{general}$. Let the intertwiner (\ref{eq:intertwiner}) be of the form
\begin{align}
	\iota_{\frac{n}{2}} : j_1 \otimes j_1 \otimes ... \otimes j_n \otimes j_n \otimes j_{n+1} \otimes j_{n+2} \to 0,
\end{align}
where the spins $j_1, ..., j_n$ appear double for they are corresponding to the holonomies with reflections, whereas $j_{n+1}$ and $j_{n+2}$ represent the holonomies which lie on the symmetry axis and hence only appear once. Without loss of generality, we can ignore $j_{n+1}$ and $j_{n+2}$ in the following considerations, as they do not contribute to a possibly resulting sign. We can now expand the intertwiner $\iota_{\frac{n}{2}}$ into a linear combination of intertwiners of 3-valent vertices, where we can freely choose the coupling scheme \cite{Wigner1993}. Consider the following expansion
\begin{widetext}
	\begin{equation}
		\underset{j_1 \qquad \qquad \qquad j_1}{\stackrel{j_n \qquad \qquad \qquad j_n}{\leftidx{}{\includegraphics[valign=c, scale=0.1]{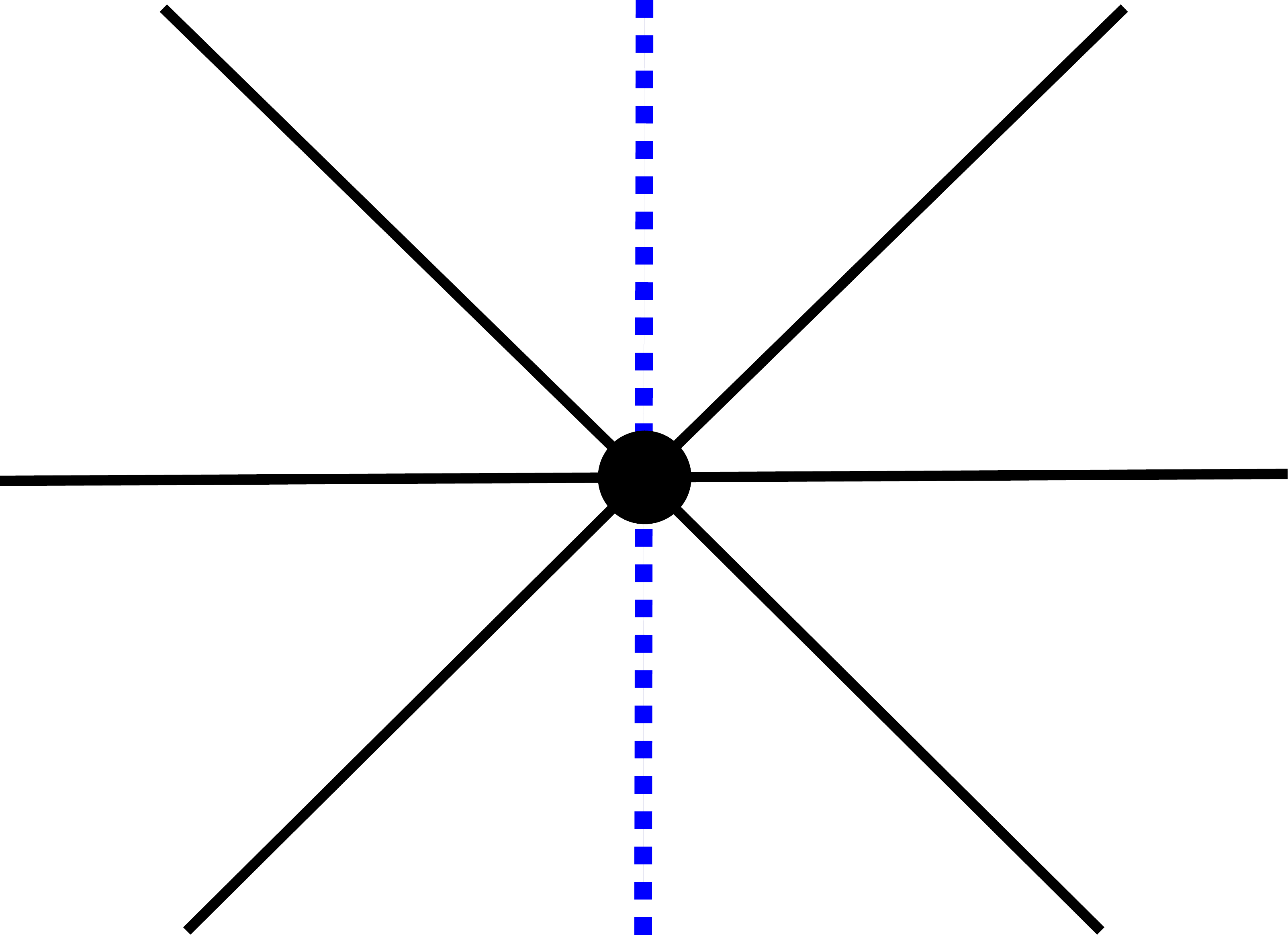}}{}}} 
		\quad \in \Spann \left\{
		\adjustbox{valign = c}{
			\begin{overpic}[scale = 0.1]{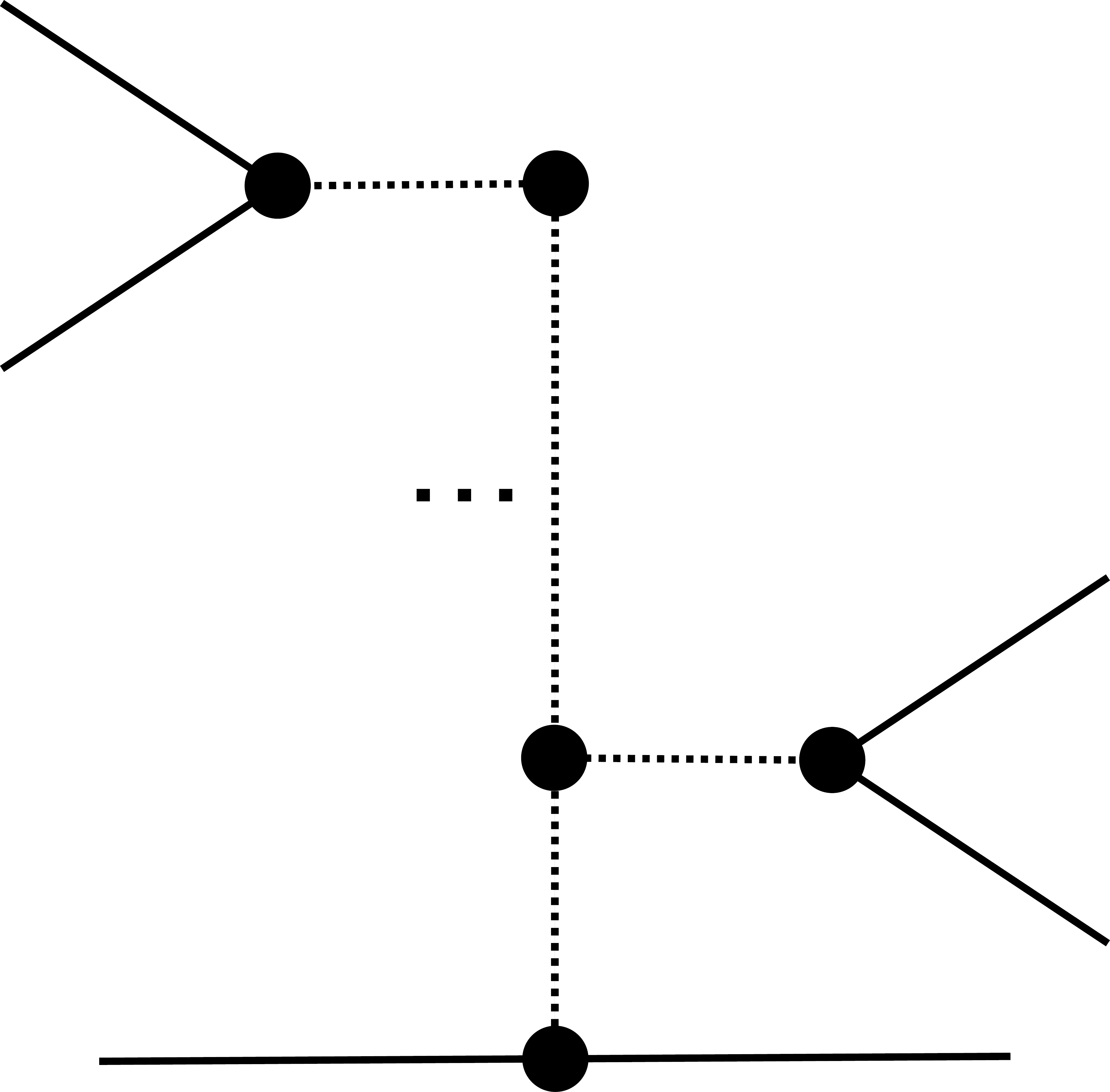}
				\put(10, 95){\scriptsize $j_n$}
				\put(10, 65){\scriptsize $j_n$}
				\put(20, -5){\scriptsize $j_1$}
				\put(75, -5){\scriptsize $j_1$}
				\put(85, 12){\scriptsize $j_2$}
				\put(85, 45){\scriptsize $j_2$}
				\put(52, 15){\scriptsize $k_1$}
				\put(52, 40){\scriptsize $k_2$}
				\put(65, 22){\scriptsize $l_2$}
				\put(35, 85){\scriptsize $l_n$}
			\end{overpic}
		}, \text{ with } \quad 
		\begin{matrix}
			k_1 = l_1, \\
			l_i \in \{0, ..., 2j_i\} \\
			k_i \in \{|l_i - k_{i-1}|, ..., l_i + k_{i-1}\} \\
			k_{n-1} = l_n
		\end{matrix}
		\right\},
		\label{eq:intertwiner_basis}
	\end{equation}
\end{widetext}
where the black dashed lines depict SU(2) representations without simultaneously describing a parallel transport in $\Sigma$, i.e. its beginning- and endpoint coincide. Note that the spin quantum numbers $l_1, ..., l_n$ as well as the $k_1, ..., k_{n-1}$ are always integers. The intertwiners of the 3-valent vertices represent the building blocks of (\ref{eq:intertwiner_basis}) and are described by the Wigner 3j-symbols \cite{Wigner1993, Regge:1958fs}, which satisfy a number of symmetry identities, in particular
\begin{align}
	\begin{pmatrix}
		j & j & l \\
		m_1 & m_2 & m_3
	\end{pmatrix}
	= (-1)^{2j + l} \begin{pmatrix}
		j & j & l \\
		m_2 & m_1 & m_3
	\end{pmatrix}.
\end{align}
This means that under the exchange of two identical spins under $\phi_\text{general}$, we gather a sign from the $i$-th branch if $2j_i$ and $l_i$ are \textit{not} both odd or both even. Consequently, all the contributions are multiplied. Obviously, this is again not a very strong statement, as we can produce arbitrary signs within the span of (\ref{eq:intertwiner_basis}). 

This indefiniteness can be illustrated with the following example
\begin{align}
	\adjustbox{valign = c}{
		\begin{overpic}[scale = 0.1]{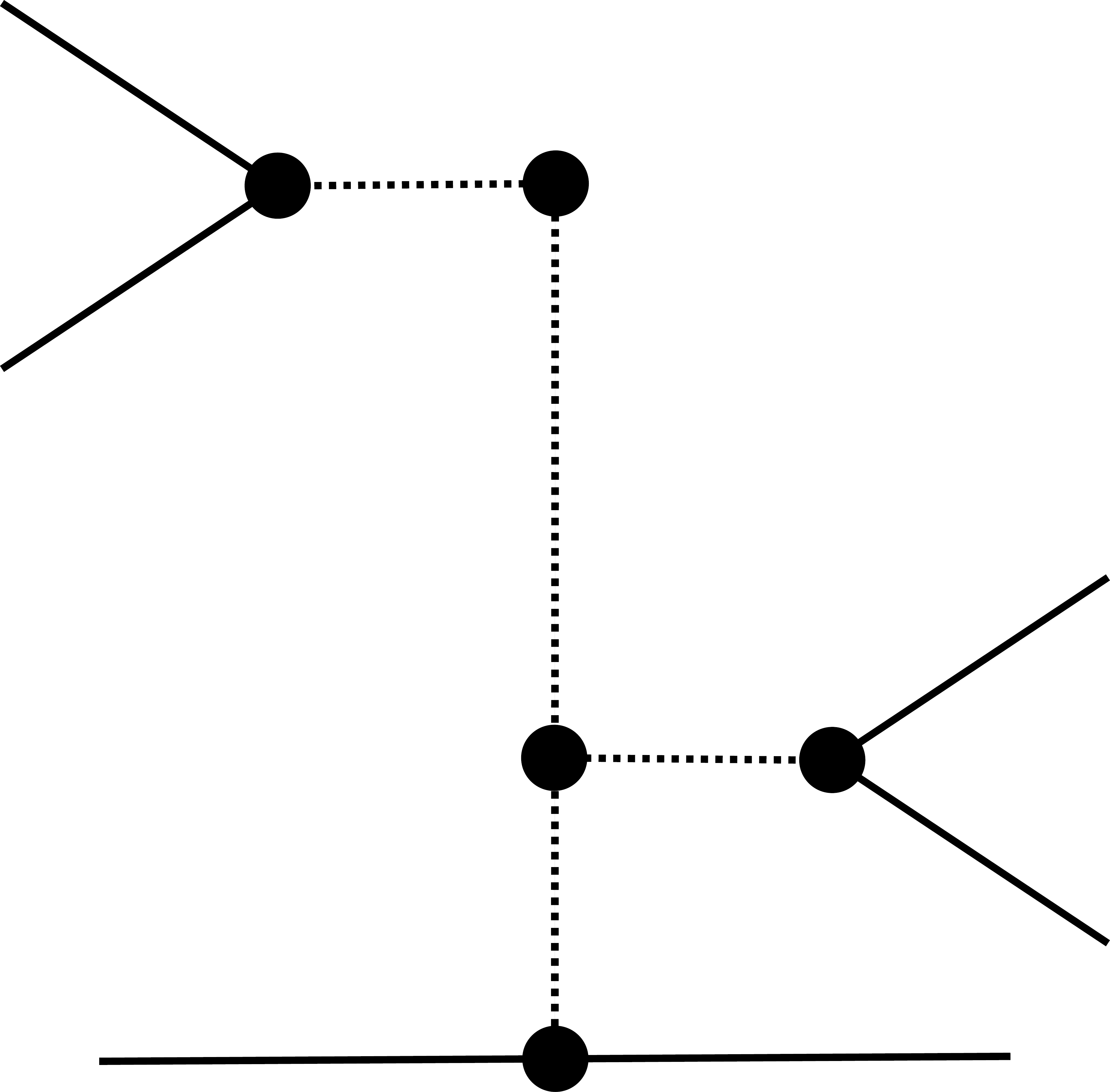}
			\put(10, 95){\scriptsize $\frac{1}{2}$}
			\put(10, 65){\scriptsize $\frac{1}{2}$}
			\put(20, -5){\scriptsize $\frac{1}{2}$}
			\put(75, -5){\scriptsize $\frac{1}{2}$}
			\put(85, 12){\scriptsize $\frac{1}{2}$}
			\put(85, 45){\scriptsize $\frac{1}{2}$}
			\put(52, 15){\scriptsize $l_1$}
			\put(52, 40){\scriptsize $k_2$}
			\put(65, 22){\scriptsize $l_2$}
			\put(35, 85){\scriptsize $l_3$}
		\end{overpic}
	}, \text{ with } \quad 
	\begin{matrix}
		k_1 = l_1, \\
		l_i \in \{0, 1\} \\
		k_{2} = l_3
	\end{matrix},
	\label{eq:counter_example_sst}
\end{align}
with $n = 3$ and $j_1 = j_2 = j_3 = \frac{1}{2}$, which is the case of a 6-valent vertex with six spin $\frac{1}{2}$ holonomies. It holds that $2j_i = 1$ for all $i = 1, 2, 3$. Hence, the sign depends on the choice of the $l_i$. However, we can choose $l_1 = l_2 = l_3 = 1$ on the one hand, and $l_1 = l_2 = 1$, $l_3 = 0$ on the other hand. The first intertwiner will not gather a phase when applying $\hat{U}_{\phi_\text{general}}$, whereas the second intertwiner gathers a sign $-1$ when doing so.

Given a generic spin network state like (\ref{eq:general_symmetric}) and an intertwiner intersecting the symmetry axis together with its expansion into the basis (\ref{eq:intertwiner_basis}), we can tell from the statistics of the particles $\theta$ which components will survive or vanish after group averaging. Specifically, fermions will let those components survive which gather a total minus sign and bosons will let those components survive which are invariant under the diffeomorphism $\phi_\text{general}$.

If we would extend the group of diffeomorphisms to include also diffeomorphisms which are smooth except at a finite number of points suggested by \cite{Fairbairn:2004qe}, the group of graph symmetries of the spin network state (\ref{eq:counter_example_sst}) is also significantly extended. It is possible to exchange any two of the edges at the vertex (\ref{eq:counter_example_sst}), for instance. However, this would generate both signs for $l_1 = l_2 = 1$ and $l_3 = 0$, i.e. the state does not survive the group averaging irrespective of the statistics of the matter fields. Apparently, this extended diffeomorphism group defines another theory and makes clear that the initial choice of the symmetry group is crucial for the analysis of the kinematical Hilbert space.

\section{An Alternative Spin 1 Action}
\label{app:U1_action}
In section \ref{subsec:Integ}, we discuss a spin 1 particle described by a tensor with an SU(2) structure $\phi_I$. If we were to consider a U(1) Yang-Mills theory\footnote{The theory is, however, not U(1) invariant, since we include a mass term.} instead, we might identify the connection 1-form $\elmagA_\mu$ with $\phi$ in the following way
\begin{align}
	\elmagA_\mu = e_\mu^I \phi_I.
\end{align} 
This transformation does not describe a symplectomorphism, since the symplectic structure is not conserved
\begin{align}
	\elmagE^\mu \dot{\elmagA}_\mu = e^\mu_I \pi^I \left( \dot{e}_\mu^J \phi_J + e_\mu^J \dot{\phi}_J \right) = \pi^I \dot{\phi}_I + e^\mu_I \dot{e}_\mu^J \pi^I \phi_J,
	\label{eq:elmag_symplectic}
\end{align}
where $\pi^I$ denotes the conjugate momentum to $\phi_I$ and $\elmagE^\mu$ the conjugate momentum to $\elmagA_\mu$. The first term in (\ref{eq:elmag_symplectic}) is the desired one but the second one is odd. Note that in flat spacetime where we require $e_\mu^I = \delta_\mu^I$, the transformation is indeed a symplectomorphism. In (\ref{eq:elmag_symplectic}) we used the identity
\begin{align}
	\elmagE^\mu = \frac{\partial L}{\partial \dot{\elmagA}_\mu} = \frac{\partial L}{\partial \dot{\phi}_I} \frac{\partial \dot{\phi}_I}{\partial \dot{\elmagA}_\mu} = \pi^I e^\mu_I.
\end{align}
This brings us to the conclusion that the two theories to describe a spin 1 field are not equivalent. Since the Yang-Mills connection $\elmagA_\mu$ does not admit an SU(2) structure, there is no interaction term including $\elmagA$ and $\omega$, albeit it is there for the action for the spin 1 field $\phi$. The matter field $\elmagA$ hence decouples from the SU(2) holonomies, which appear in matter-free loop quantum gravity.

\section{Constraints of the Spin 1 Action}
\label{app:constraints}
The action (\ref{eq:spin1_action}) resembles very much the Klein-Gordon action but now including a nontrivial SU(2) interaction. We want to mention some technical details and hurdles which arise when studying the classical constraints. As in the matter-free theory, $\omega_0$ still has to be reformulated to yield the Lagrange multiplier for the Gauss constraint. The conjugate momentum $\pi$ reads
\begin{align}
	\pi^I &= \sqrt{\det(q)} \left( \frac{1}{N} \nabla_0 \phi^I - \frac{N^a}{N} \nabla_a \phi^I + \eta^{0I} \nabla^K \phi_K \right).
	\label{eq:spin1_momentum}
\end{align}
We can read off that $\pi^0$ is independent of the time derivative of $\phi_0$ since the first and part of the third term in (\ref{eq:spin1_momentum}) cancel. We get
\begin{align}
	\pi^0 = - \sqrt{\det(q)} e^a_i \nabla_a \phi^i.
\end{align}
As a consequence, we get an additional constraint
\begin{align}
	f(X) := \int_\Sigma d^3x X f = \int_\Sigma d^3x X \left( \pi^0 + \sqrt{\det(q)} e^a_i \nabla_a \phi^i \right),
	\label{eq:add_constraint}
\end{align}
which we have to add to the action (\ref{eq:spin1_legendre}). $X$ here acts as the Lagrange multiplier. As we can see, the term $\nabla^I \phi_I \cdot f$ already appears in the Hamilton constraint and can therefore be left out when doing the constraint analysis. For this, we would have to calculate the Poisson brackets of $f(X)$ with the other constraints to identify possible secondary constraints.\footnote{The constraint analysis for a spin 1 particle in flat spacetime 
	yields a secondary constraint $g(Y) = \int_\Sigma d^3 x Y \left( \partial_i \pi^i - (\partial_a \partial^a - m^2) \phi_0 \right)$. Its Poisson bracket with $f(X)$ is 
	constant, 
	so $f(X)$ and $g(Y)$ form a second class pair. The constraint algebra is closed, but the variables $\pi^0$ and $\phi_0$ are determined by the solution of $f$ and $g$, namely $\pi^0 = - \partial_i \phi^i$ and $\phi_0 = - (\partial_a \partial^a - m^2)^{-1} \partial_i \pi^i$. $\phi_0$ hence is nonlocal. 
}

At last, we express the spin connection $\omega$ by the Ashtekar connection $A$. In the matter-free theory, the spin connection can be written in terms of two variables
\begin{align}
	\omega_a^{0i} &= \frac{1}{2 \gamma} \left( \leftidx{}{A}{_a^i} - \leftidx{^-}{A}{_a^i} \right) \\
	\tensor{\epsilon}{^i_{jk}} \omega_a^{jk} &= \frac{1}{2} \left( \leftidx{}{A}{_a^i} + \leftidx{^-}{A}{_a^i} \right),
\end{align}
where $A_a$ is the Ashtekar connection and $\leftidx{^-}{A}{_a}$ is non-dynamical \cite{Holst_96}. Next to the lapse function and the shift vector field, we hence also have the Lagrange multipliers $\omega_0^{0i}$, $\omega_0^{jk}$ and $\leftidx{^-}{A}{_a}$. In the vacuum theory, hence, the following constraints
\begin{align}
	\frac{\partial L}{\partial \left( \omega_0^{0i} \right)} = 0 \qquad \frac{\partial L}{\partial \left( \omega_0^{jk} \right)} = 0 \qquad \frac{\partial L}{\partial \left( \leftidx{^-}{A}{^i_a} \right)} = 0
\end{align}
hold and yield the well-known Gauss constraint next to defining relations for the Lagrange multiplier $\omega_0^{0i} = \omega_0^{0i}(A, \Gamma, N, N^a)$ and the connection $\Gamma_a^i = \Gamma_a^i (E)$ such that
\begin{align}
	\leftidx{^-}{A}{_a^i} = A_a^i - \frac{2}{\gamma} \Gamma_a^i.
\end{align}
If we include the spin 1 field, however, we get the constraints
\begin{widetext}
	\begin{align}
		\frac{\partial L}{\partial \left( \omega_0^{0i} \right)} &= - \partial_a E^a_i - \tensor{\epsilon}{^k_{li}} E^a_k \left( \frac{\gamma^2 + 1}{2 \gamma} \leftidx{^-}{A}{_a^l} - \frac{\gamma^2 - 1}{2 \gamma} A_a^l \right) + 2 \phi_{[0} \pi_{i]} = 0 \label{eq:secondclassconst1} \\
		\frac{\partial L}{\partial \left( \omega_0^{jk} \right)} &= \frac{\gamma}{2} \tensor{\epsilon}{^i_{jk}} \partial_a E^a_i - E^{a [j} A^{k]}_a + 2 \phi_{[j} \pi_{k]} = 0 \label{eq:secondclassconst2} \\
		\frac{\partial L}{\partial \left( \leftidx{^-}{A}{^i_a} \right)} &= - \frac{\gamma^2 + 1}{2 \gamma} \tensor{\epsilon}{^k_{im}} E^a_k \omega_0^{0m} + \frac{\gamma^2 + 1}{2 \gamma} \epsilon^{abc} e_{b[i|} e_{d |j]} N^d \left( A_c^j + \leftidx{^-}{A}{_c^j} \right) + \nonumber \\
		&- \frac{\gamma^2 + 1}{2 \gamma} \epsilon^{abc} \partial_b (N e_{c i}) + \frac{\gamma^2 + 1}{2 \gamma^2} \epsilon^{abc} \epsilon_{ijk} N e_b^j \left( A_c^k - \leftidx{^-}{A}{_c^k} \right) + \nonumber \\
		&+ N^a \left( 2 \tensor{\epsilon}{_i^{jk}} \phi_j \pi_k - 2 \phi_{[0} \pi_{i]} \right) - q^{ab} N \left( \tensor{\epsilon}{_i^{jk}} \phi_{j} \nabla_b \phi_{k} - \phi_{[0} \nabla_b \phi_{i]} \right) + \nonumber \\
		&- e^a_{[j|} \left( \tensor{\epsilon}{_i^{jk}} \phi_{|k]} \nabla^l \phi_l + \delta_i^j \phi_{|0]} \nabla^l \phi_l \right) - \frac{1}{2} e^a_{[j|} \left( \tensor{\epsilon}{_i^{jk}} \phi_{|k]} \nabla^0 \phi_0 + \delta_i^j \phi_{|0]} \nabla^0 \phi_0 \right) + \nonumber \\
		&+ \pi_0 e^a_{[j|} \left( \tensor{\epsilon}{_i^{jk}} \phi_{|k]} + \delta_i^j \phi_{|0]} \right) = 0. \label{eq:secondclassconst3}
	\end{align}
\end{widetext}
It becomes apparent that the split of the indices of the spin connection $\omega_a$ into $0$ and $i$ makes it troublesome to recover the desired information from (\ref{eq:secondclassconst1} - \ref{eq:secondclassconst3}). When including the spin 1 field, we only get a contribution of the $j$ component of the matter field. The $0$ component on the other hand does not appear in the Gauss constraint but rather in the defining relations for $\Gamma$ and $\omega_0^{0i}$ if we follow the same steps as in the matter-free case.

\newpage

\bibliography{literature}

\end{document}